\documentclass[12pt,preprint]{aastex}

\shorttitle{Chandra Observation of Abell 2052}
\shortauthors{Blanton, Sarazin, \& McNamara}

\slugcomment{Astrophysical Journal, in press}

\begin{document}

\title{{\it Chandra} Observation of the Cooling Flow Cluster Abell 2052}

\author{Elizabeth L. Blanton\altaffilmark{1,2},
Craig L. Sarazin\altaffilmark{1},
and Brian R. McNamara\altaffilmark{3}}

\altaffiltext{1}{Department of Astronomy, University of Virginia,
P. O. Box 3818, Charlottesville, VA 22903-0818;
eblanton@virginia.edu, sarazin@virginia.edu}

\altaffiltext{2}{{\it Chandra} Fellow}

\altaffiltext{3}{Department of Physics \& Astronomy, Ohio University,
Clippinger Labs, Athens, OH 45701;
mcnamara@helios.phy.ohiou.edu}

\begin{abstract}
We present an analysis of the {\it Chandra} X-ray observation of Abell 2052,
including large scale properties of the cluster as well as the central region
which includes the bright radio source, 3C 317.
We present temperature and abundance profiles
using both projected and deprojected spectral analyses.
The cluster shows the cooling flow signatures of excess surface
brightness above a $\beta$-model at the cluster center, and a temperature 
decline into the center of the cluster.  For Abell 2052, the temperature
drops by a factor of three from approximately 3 to 1 keV.
The heavy element abundances initially increase into the center, but
decline within 30\arcsec.
Temperature and abundance maps show that the X-ray bright shells surrounding
the radio source are the coolest and least abundant regions in the cluster.
The mass-deposition rate
in the cooling flow is $26 < \dot{M} < 42~M_{\odot}$ yr$^{-1}$.
This rate is approximately a factor of three lower than the rates found with 
previous X-ray observatories.
Based on a stellar population analysis using imaging and spectra
at wavelengths spanning the far ultraviolet to the near infrared,
we find a star formation rate of $0.6~M_{\odot}$ yr$^{-1}$ within a 
$3\arcsec$ radius of the nucleus of the central cluster galaxy.
Total and gas mass profiles for the cluster are also determined.
We investigate additional sources of pressure in the X-ray holes formed by
the radio source, and limit the temperature of any hot, diffuse, thermal
component which provides the bulk of the pressure in the holes to 
$kT \ga 20$ keV.  
We calculate the magnetic field
in the bright-shell region and find $B \approx 11~\mu$G.  The magnetic 
pressure in the cluster center is significantly lower than the gas pressure.
The current luminosity of the central AGN is $L_X = 7.9 \times 10^{41}$
erg s$^{-1}$, and its spectrum is well-fitted by a power-law model with
no excess absorption above the Galactic value.
The energy output from 
several radio outbursts, occurring episodically over the lifetime of the 
cluster, may be sufficient to offset the cooling flow near the center.
\end{abstract}

\keywords{
cooling flows ---
galaxies: clusters: general ---
galaxies: clusters: individual (Abell 2052) ---
intergalactic medium ---
radio continuum: galaxies ---
X-rays: galaxies: clusters
}

\section{Introduction}

Cooling flows are expected to occur in the centers of clusters of galaxies
where the gas density is high and the cooling time is short
(see Fabian [1994] for a review).
When the gas
cooling time is shorter than the age of the cluster (or the time since its
last major merger), the gas at the center cools and outer gas flows in to
maintain hydrostatic equilibrium.
With the stream of new data coming from the {\it Chandra} and {\it XMM-Newton}
X-ray Observatories, our picture of cooling flow clusters has changed
dramatically.

The vast majority of cooling flow clusters contain powerful radio sources
associated with central cD galaxies.
High-resolution imaging results from {\it Chandra} revealed that these
radio sources have a profound effect on the intracluster medium (ICM) --
the radio lobes displace the X-ray emitting gas creating X-ray deficient
``holes'' or ``bubbles.''  Some evidence of this was found with {\it ROSAT} 
observations of Perseus (B\"ohringer et al.\ 1993), Abell 4059 
(Huang \& Sarazin 1998), and Abell 2052 (Rizza et al.\ 2000).  
The {\it Chandra} 
high-resolution observations have found many more 
cases and allow us to study the physics of the interaction in much more 
detail (i.e. Hydra A, McNamara et al.\ 2000; Perseus, Fabian et al.\ 2000;
Abell 2052, Blanton et al.\ 2001; Abell 2597, McNamara et al.\ 2001;  Abell 496,
Dupke \& White, 2001; MKW3s, Mazzotta et al.\ 2001; RBS797, Schindler et al.\
2001; Abell 2199, Johnstone et al.\ 2002; Abell 4059, Heinz et al.\ 2002;
Virgo, Young et al.\ 2002;
Centaurus, Sanders \& Fabian 2002; Cygnus A, Smith et al.\ 2002).

A long-standing problem with cooling flow models has been that the mass of
gas measured to be cooling from X-ray temperatures, based on 
surface-brightness and spectral
studies with {\it Einstein}, {\it ROSAT}, and {\it ASCA},
has not shown up in sufficient quantities at cooler temperatures.
While significant star formation has been found in cD galaxies at the
centers of cooling flows, these rates are only about $1-10\%$ of that
predicted from the gas inflow rates determined by
morphological X-ray studies (McNamara 1997).  Moderate resolution spectra
from {\it Chandra} have been used to recalculate the mass inflow rates in
cooling flows (e.g. David et al.\ 2001), and these rates are typically 
much lower than those inferred from the surface-brightness studies, and more 
in line with what is
seen optically.  High-resolution spectroscopy with {\it XMM-Newton} provided
direct evidence that gas was cooling in these clusters, but very large 
masses of gas (hundreds to thousands of solar masses) were seemingly 
cooling over only
a limited range of temperatures.  Emission lines such as Fe XVII expected 
from gas cooling below approximately 2 keV were not detected
(Kaastra et al.\ 2001;  
Peterson et al.\ 2001;
Tamura et al.\ 2001).  
The {\it XMM-Newton} observations of Abell 1835 (Peterson et al.\ 2001),
for example, limited the amount of gas cooling below 2 keV to 
$\dot{M} < 200 \, M_{\odot}$ yr$^{-1}$, compared to the value of
$\dot{M} \approx 2000 \, M_{\odot}$ yr$^{-1}$ determined from {\it ROSAT} and
{\it ASCA} data (Allen et al.\ 1996).
Several possible solutions have been proposed for the lack of cool 
($kT \la 2$ keV) X-ray gas seen in the new observations
(Fabian et al.\ 2001;
Peterson et al.\ 2001).
These include mixing, heating, inhomogeneous
abundances, and differential absorption.
Heating of the gas by a central radio source is also discussed by
B\"ohringer et al.\ (2002) and Churazov et al.\ (2002). 

In Blanton et al.\ (2001; hereafter Paper I), we examined the interaction 
of the central
radio source (3C 317) and the X-ray emitting gas in the cooling flow 
cluster Abell 2052 at a redshift of $z=0.0348$.  We found one of the clearest cases of the radio source
displacing, and in turn being confined by, the X-ray gas.
X-ray deficient holes were found that corresponded with the radio emission, and
the holes were surrounded by bright shells of dense, X-ray emitting gas.
As has
been found with other clusters, such as Perseus (Fabian et al.\ 2000), these
shells were cool and showed no evidence of current strong shocks.  In
Soker, Blanton, \& Sarazin (2002), we found that the morphology was 
well-explained by weak shocks occurring in the past, and strong shocks were
generally ruled out.  Therefore, strong-shock 
heating from the radio source as proposed by Heinz, Reynolds, \& Begelman
(1998) and Rizza et al.\ (2000) is probably not a good explanation
for the fate of the missing cool gas, although long-timescale energy input
from a radio source can still contribute to heating.

In this paper, our focus is a spatially larger scale study of 
the cooling flow in Abell 2052, as well as further analysis on 
the cluster center and the interaction of the radio source with the 
intracluster medium (ICM).
We assume $H_{\circ}=50$ km s$^{-1}$ Mpc$^{-1}$ and $q_{\circ}=0.5$ 
(1\arcsec\ = 0.95 kpc at $z=0.0348$) throughout.

\section{Observation and Data Reduction}

Abell 2052 was observed with {\it Chandra} on 2000 September 3 for a total of
36,754 seconds.  The observation was taken so that the center of the cluster
would fall near the aimpoint of the ACIS-S3 CCD.  In addition to the S3, data
were received from the ACIS I2, I3, S1, S2, and S4 CCDs.  In the analysis that
follows, we use data from the S3 only.  The events were telemetered in Faint 
mode, the data were collected with frame times of 3.2 seconds, and the CCD
temperature was $-120$ C.  Only events with {\it ASCA} grades of 
0,2,3,4, and 6 were included.  
Unless otherwise noted, background was taken from the blank sky observations 
collected by
M. Markevitch\footnote{See http://cxc.harvard.edu/contrib/maxim/acisbg/}.

We used the {\it Chandra} data analysis package CIAO v2.1 for the
data reductions.
Bad pixels, bad columns, and columns next to bad columns 
and node boundaries were excluded.  For the analysis, event PI values 
and photon energies were
determined using the acisD2000-08-12gainN0003.fits gain file.
The data were searched for background flares and none were found.  A small
period of bad aspect was found, and as a result 132 seconds of data were
excluded, leaving a total exposure of 36,622 seconds.  Recently, degradation
in the quantum efficiency of the ACIS detector at low energies has been
discovered which affects the results of spectral fitting.  
The degradation is probably the result of molecular gas buildup
on the CCD chips and/or the optical blocking filter.  We have used the
corrarf.f program supplied by the Chandra X-ray Center as of 2002 August to
correct our effective area files (arfs) for this effect before performing
our spectral fits.  While this 
correction gives a significantly more realistic picture of the data at 
low energies than
no correction, it is still uncertain, and values of absorption, which depend
most heavily on the low energy region, should be viewed as uncertain at this
point.  All spectral fitting was done with XSPEC v11.1.0.

\section{X-ray Image}

\begin{figure}
\plotone{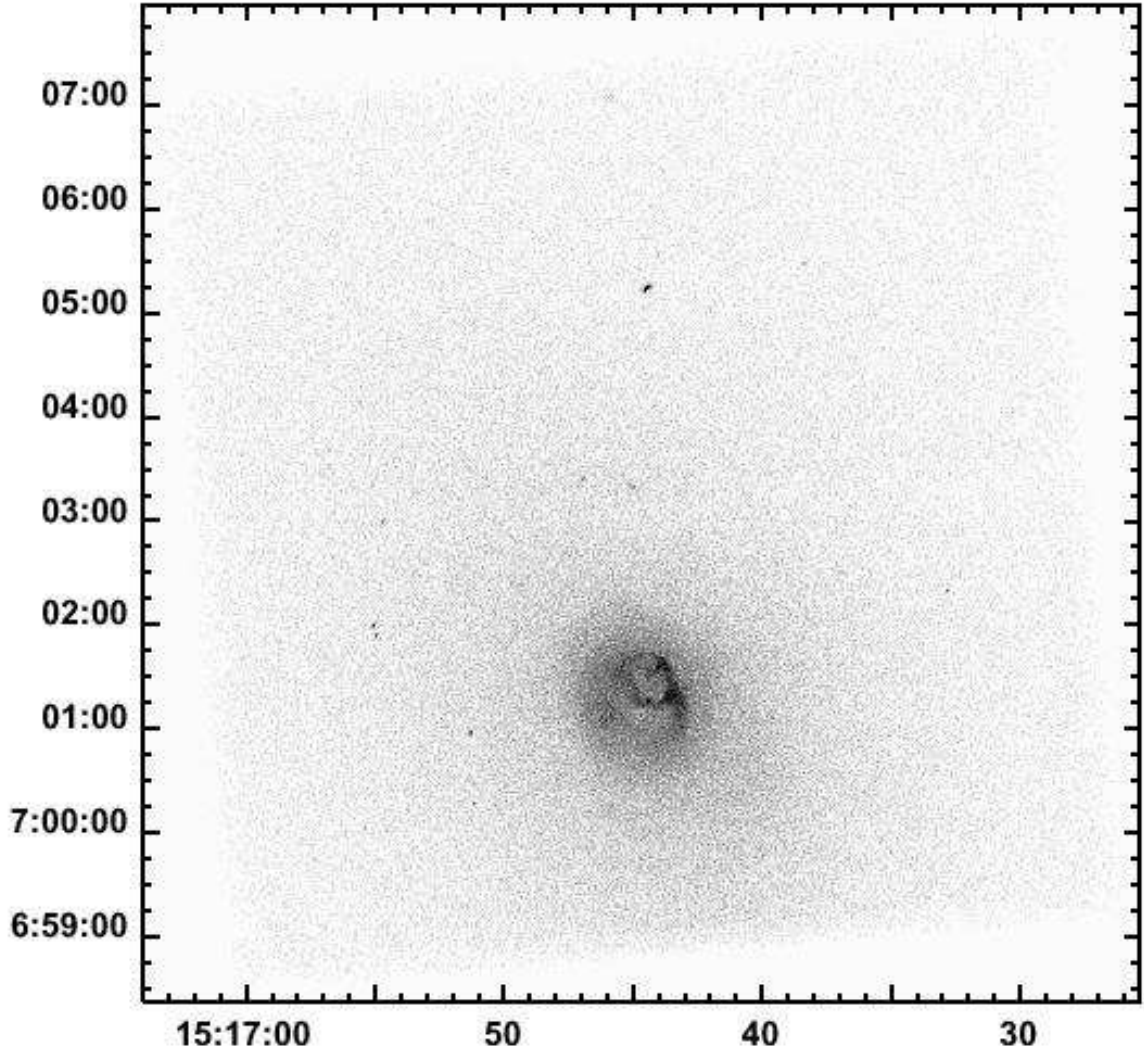}
\caption{Unsmoothed 0.3 -- 10.0 keV band image of Abell 2052 from 
the {\it Chandra} ACIS-S3, uncorrected for background or exposure.
Shells and voids are apparent in the inner
regions of the cluster, as well as a central point source corresponding
to the AGN (3C 317).
\label{fig:xrayall}}
\end{figure}

\begin{figure}
\plotone{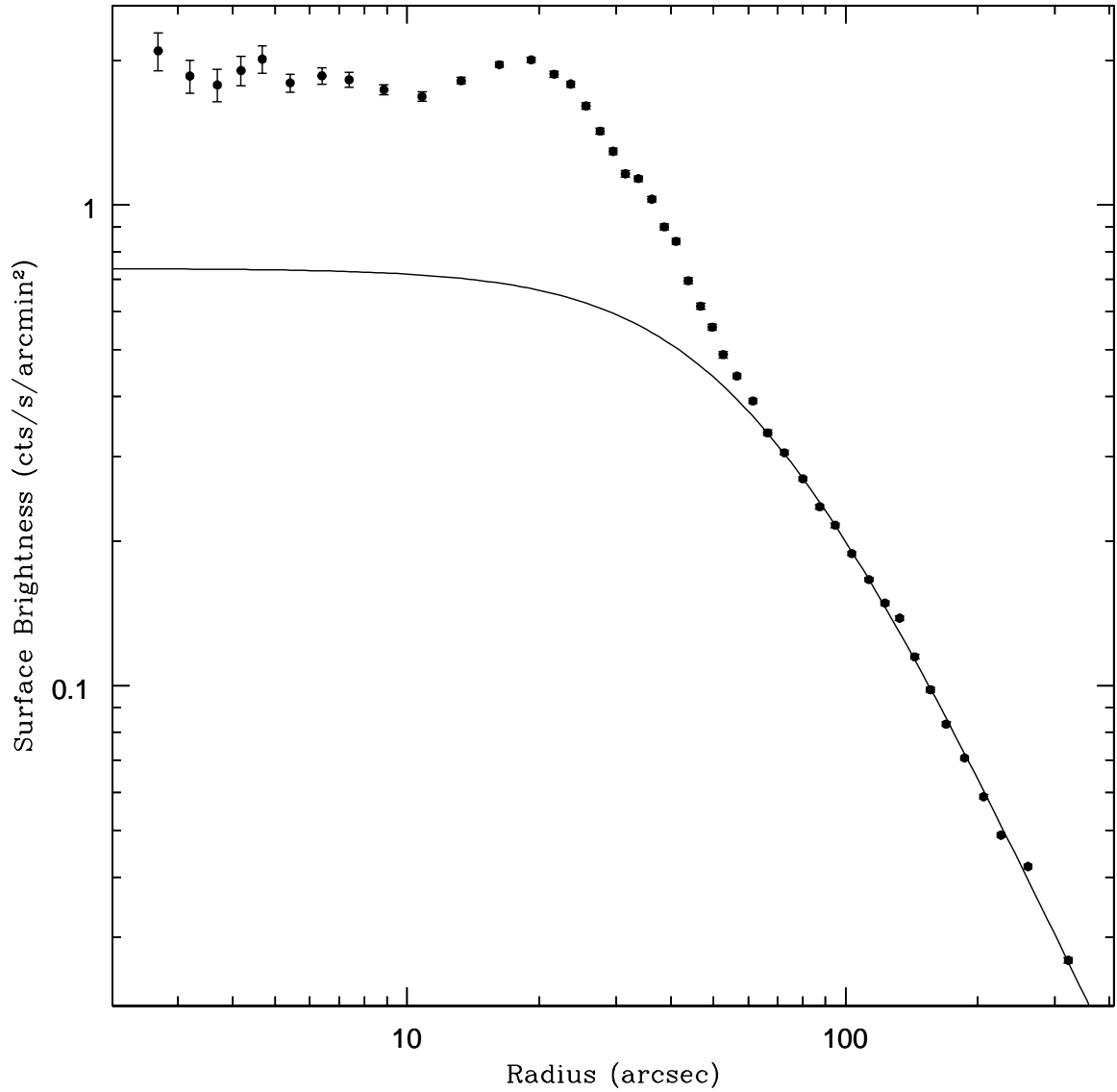}
\caption{X-ray surface brightness profile from the {\it Chandra} observation
of Abell 2052.
The radius is measured from the position of the central AGN.
The error bars are at the $1-\sigma$ level.
The solid line is the fit of a single $\beta$-model including only points
beyond 70 arcsec in the fit.  The surface brightness profile shows an
excess in the center above the $\beta$-model which is characteristic of a 
cooling flow.
\label{fig:surfbr}}
\end{figure}

The entire unsmoothed ACIS-S3 image in the 0.3 -- 10.0 keV band is shown 
in 
Figure~\ref{fig:xrayall}.  The image has not been corrected for background
or exposure.  There is a strong increase in brightness at the 
center of the cluster, typical of a cooling flow.
A plot of the surface
brightness profile is shown in Figure~\ref{fig:surfbr},
corrected for exposure and background.
The plot shows the fit (solid line) of a $\beta$-model, using only points
with radii greater than 70 arcsec in the fit.  The model
has a best-fitting $\beta$-value of $\beta = 0.49$, and a core radius of 
59 arcsec.
Fitting the entire profile with a $\beta$-model gave a very poor fit,
as judged by the $\chi^2/$d.o.f.\ value, and the largest residuals were
at low radii.  We therefore eliminated points in steps of 10 arcsec from 
the center and fitted a series of $\beta$-models until a minimum
in $\chi^2/$d.o.f.\ was reached
($\chi^2/$d.o.f.\ was no longer improved by eliminating more points).
This minimum was reached at 70 arcsec from the center.  The 
surface brightness profile shows the signature of a cooling
flow --- excess emission in the cluster center relative to a $\beta$-model.
In Figure~\ref{fig:xrayall}, substructure is apparent in the inner regions 
of the gas distribution, including two shells of 
emission to the north and south of the cluster center, surrounding two 
``holes'' --- regions of lower surface brightness.
These regions result from the central radio source displacing and compressing 
the X-ray gas, and were discussed in detail in Paper I.

\begin{figure}
\plotone{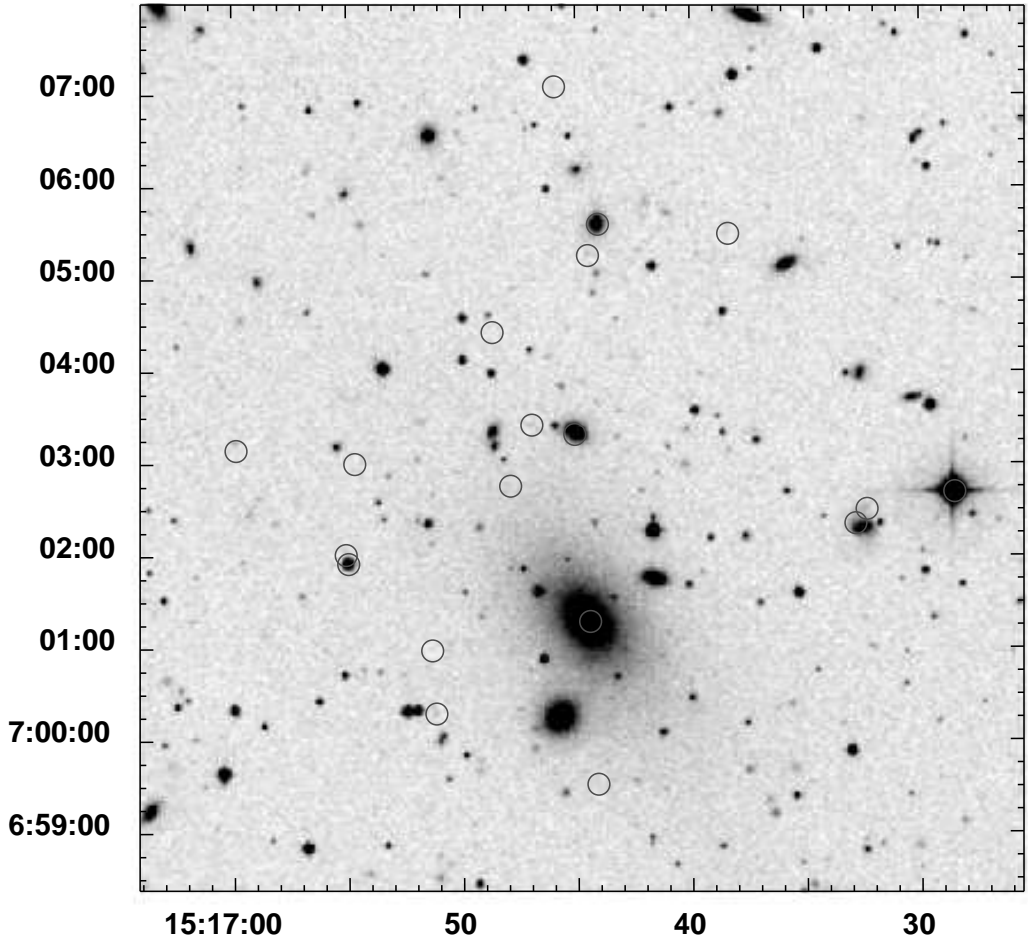}
\caption{Optical image from the Digitized Sky Survey, trimmed to match the
field of few shown in Figure~\ref{fig:xrayall}.  X-ray point sources 
detected in the {\it Chandra} ACIS-S3 observation are marked with circles.
\label{fig:dss}}
\end{figure}

In addition to the diffuse emission, a number of discrete sources are seen
on the image.  We used a wavelet detection algorithm (WAVDETECT in CIAO) to 
detect individual sources.  The source detection threshold was set at 
$10^{-6}$, implying that $\la 1$ false source (due to a statistical 
fluctuation) would be detected within the area of the S3 image.  Sources were
visually confirmed on the X-ray image, and a few low-level detections at the
edges of the field in regions of low exposure were removed.
Using this method, we found nineteen individual sources.
One of these corresponds to the
AGN, 3C 317, in the central cD,
two others correspond to other galaxies in the cluster (CGCG 049-091 and
PGC 054524),
while another two seem to be associated with stellar sources (BD+07 2929 and
an uncataloged optical source).
For these five objects, all of the
X-ray positions agree within approximately 1 arcsec with the USNO optical
positions, with the exception of the AGN --- it is offset
$1\farcs7$ from the
USNO position for the cD galaxy.  However, the AGN's X-ray position is offset
only $0\farcs7$ from the radio core position given in Zhao et al.\ (1993).
An image from 
the Digitized Sky Survey (DSS), trimmed to show the same field-of-view as the 
ACIS-S3 and with the nineteen X-ray sources marked, is displayed in 
Figure~\ref{fig:dss}.

\section{Spectral Analysis}

\subsection{Total Spectrum} \label{sec:totspec}
In an effort to understand large-scale features of the ICM in Abell 2052, 
including the mass-deposition rate of the cooling flow, we extracted a 
spectrum from a circular region with a radius of 130 kpc (136.8 arcsec),
excluding the AGN and other point sources.
This was the largest region that would fit wholly on the ACIS-S3, centered 
on the cluster, without going over the edges of the chip.  The spectrum
contained a total of 156,300 background-subtracted counts in the 0.7-10.0 keV
range, with the background taken from the blank sky observations of 
Markevitch\footnotemark[4].
In order to more realistically represent the errors and
the $\chi^2$ determinations from the spectral fits, we added a systematic
error of $2\%$ when fitting the spectrum in XSPEC.  We exclude the region from
1.8--2.1 keV in our fits because of the uncertainty of the response in this
`iridium edge' region.
All of the spectral fits were done twice, once with absorption 
fixed to the Galactic value of $2.85 \times 10^{20}$ cm$^{-2}$
(Dickey \& Lockman 1990), and once with the absorption left as a free
parameter.  The fits are summarized in Table~\ref{tab:tot}.

\subsubsection{Fixed Galactic Absorption}

In this subsection, we describe the spectral fits that included absorption
fixed to the Galactic value.
We first attempted to fit the data with a single-temperature MEKAL model
(1-MEKAL in Table~\ref{tab:tot}).
This gave
$\chi^2/$d.o.f.\ $ = 676.3/475 = 1.42$,
a temperature of $kT = 2.72^{+0.035}_{-0.025}$ keV and an abundance of 
$0.53^{+0.024}_{-0.028}$ solar.
The spectrum was better described by a two-MEKAL (2-MEKAL in 
Table~\ref{tab:tot}) model 
($\chi^2/$d.o.f.\ $ = 560.9/473 = 1.19$),
with temperatures for
the two components of $kT_{\rm low} = 0.77^{+0.065}_{-0.053}$ keV and 
$kT_{\rm high} = 2.87^{+0.038}_{-0.022}$ keV.
For this fit, the abundances of the two thermal components, which were
assumed to be identical, were $0.64^{+0.042}_{-0.032}$ solar.

\begin{figure}
\vskip6.5truein
\includegraphics{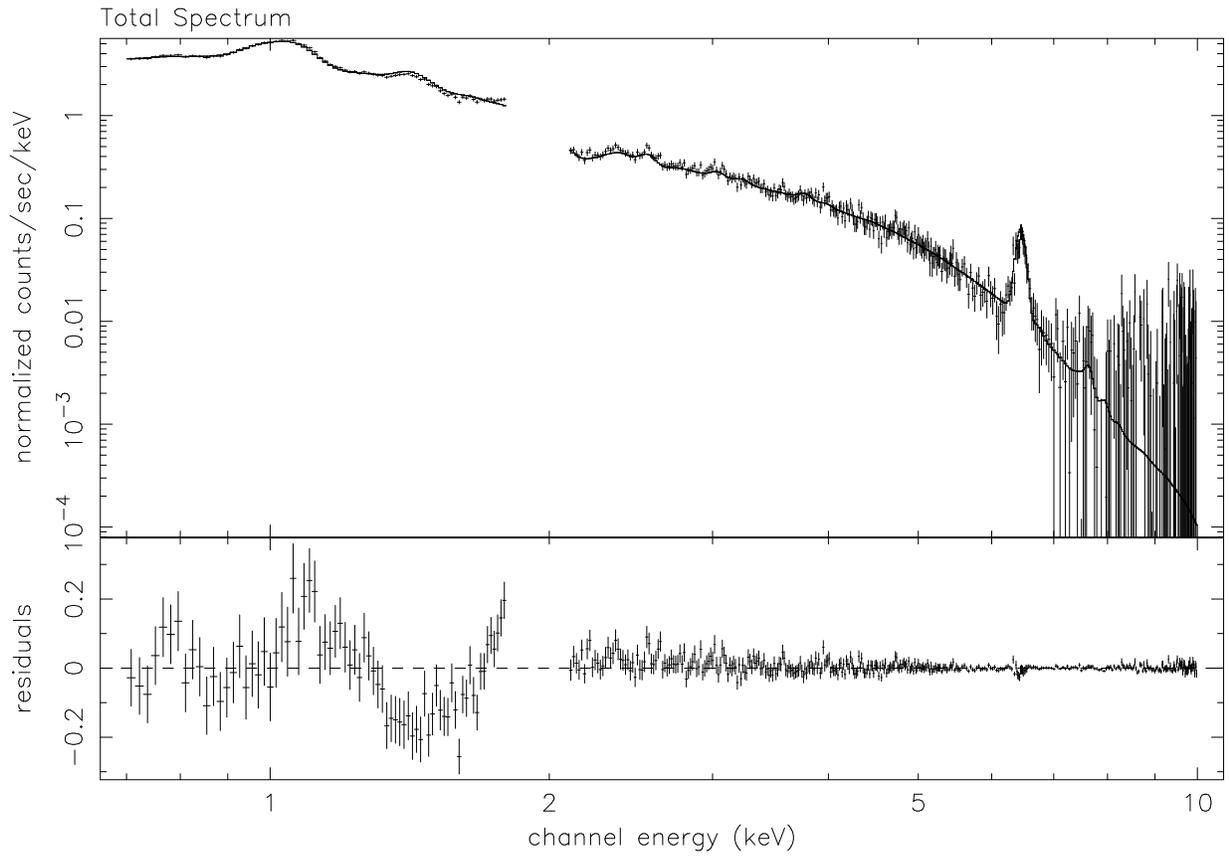}
\caption{Spectrum extracted from a circular region with a radius of
137\arcsec\ centered on the AGN.  The model shown includes Galactic 
absorption, a cooling flow, and a MEKAL component.
\label{fig:total}}
\end{figure}

We also fitted the spectrum with a model (MKCFLOW+MEKAL in 
Table~\ref{tab:tot}) combining a cooling flow component
(MKCFLOW) with a MEKAL component (to account for the gas in the 
outer regions).
We set the upper temperature limit of the MKCFLOW model to the
temperature of the MEKAL model, as would be expected if the gas cooled
from ambient ICM.  If we also set the abundances to be equal, this gave
$kT_{\rm low} = 0.43^{+0.12}_{-0.43}$ keV, 
$kT_{\rm high} = 2.95^{+0.069}_{-0.021}$ keV, and
an abundance of $0.64^{+0.027}_{-0.047}$ solar, with
$\chi^2/$d.o.f.\ $ = 554.6/473 = 1.17$.  The low temperature with the errors,
is consistent 
with the gas cooling to very low values, below where it would be detected in
the X-ray. 
A plot of this model fitted to the spectrum is shown in Figure~\ref{fig:total}.
The mass-deposition rate given from this fit is 
$\dot{M} = 32^{+4}_{-4}~ M_{\odot}$ yr$^{-1}$, which
is lower than the values measured from 
{\it Einstein} (White, Jones, \& Forman 1997),  
{\it ROSAT} (Peres et al.\ 1998), and 
{\it ASCA} (White 2000) of approximately $90 - 120 M_{\odot}$ yr$^{-1}$ for
a similar region of the cluster.
If we fix the low 
temperature component of the cooling flow to a low value 
($kT_{\rm low} = 0.001$ keV), as would be expected from the standard 
cooling flow model, the fit is just as good
($\chi^2/$d.o.f.\ $ = 554.9/474 = 1.17$), and the 
mass-deposition
rate is $\dot{M} = 28^{+4}_{-2}~ M_{\odot}$ yr$^{-1}$ (2-MEKAL-abs1 in
Table~\ref{tab:tot}).
The mass of gas cooling down to very low temperatures (if
absorption is set to the Galactic value) is 
$26 < M_{\odot} < 36$ yr$^{-1}$.

Finally, we used a model with individually-varying elemental abundances 
(VMEKAL) to fit the spectrum.
We first fitted the spectrum with all of the
elemental abundances free;
this produced a good fit, but most of the abundances were poorly
constrained.
Then, based on the results of this fit, we 
grouped together the elements that showed similar abundances in order to 
constrain them better.
The grouped elements were assumed to have the same abundance relative to
solar.
We coupled together C, N, and O; Ne and Na; Mg and Al;
and Ar, Ca, Fe, and Ni.  Si and S were left ungrouped, and He was fixed at 
the solar value.
We included the 1.8--2.1 keV region in this fit, because without it Si
(which has a prominent line at 1.8 keV)
had a poorly constrained abundance.
After fitting, we then ignored the 1.8--2.1 keV region in order to 
quote the $\chi^2$ for the same energy region as the other fits.
The best fit gave
$\chi^2/$d.o.f.\ $ = 541.99/470 = 1.15$,
and a temperature of $kT = 2.70^{+0.03}_{-0.05}$ keV.
Those elements that had strong lines in the energy range of the
fitted-spectrum (0.7-10.0 keV) that were well-fitted by the model were
Si with an abundance of $0.63^{+0.08}_{-0.07}$ solar, 
S with an abundance of $0.78^{+0.12}_{-0.10}$ solar, and Fe with 
an abundance of 
$0.59^{+0.07}_{-0.04}$ solar.
If we include the 0.5--0.7 keV range in the fit, we can constrain the 
O abundance to $0.60^{+0.11}_{-0.08}$ solar, and the values for the other
elements when fitted in this expanded range are consistent with the values
found when fitting the 0.7--10.0 keV range.

\subsubsection{Free Absorption}

We followed the same sequence of models described above, but the absorption
was allowed to vary freely.  In all of the free-absorption fits other
than the single temperature MEKAL model, the 
best-fitting model resulted in absorption that was consistent with
the Galactic value.
The single-temperature MEKAL model (1-MEKAL-abs in Table~\ref{tab:tot}) gave
$\chi^2/$d.o.f.\ $ = 599.9/474 = 1.27$,
$N_{H} = (0.00^{+0.24}_{-0.00}) \times 10^{20}$ cm$^{-2}$, 
a temperature of $kT = 2.86^{+0.029}_{-0.025}$ keV and an abundance of 
$0.60^{+0.022}_{-0.038}$ solar.
As in the fixed-absorption case, the spectrum was better 
described by a two-MEKAL model 
($\chi^2/$d.o.f.\ $ = 560.8/472 = 1.19$; 2-MEKAL-abs in Table~\ref{tab:tot}),
and had $N_{H} = (2.34^{+0.97}_{-0.40}) \times 10^{20}$ cm$^{-2}$, 
temperatures for
the two components of $kT_{\rm low} = 0.81^{+0.068}_{-0.048}$ keV and 
$kT_{\rm high} = 2.90^{+0.040}_{-0.028}$ keV,
and an abundance of $0.66^{+0.025}_{-0.043}$ solar.

The best-fitting free-absorption model combining a cooling flow component
(MKCFLOW) with a MEKAL component (MKCFLOW+MEKAL-abs1 in Table~\ref{tab:tot})
was similar to the
fixed-absorption case.  The best-fit achieved was consistent with the
gas cooling all of the way down to very low temperatures (outside of 
X-ray detection) with
$kT_{\rm low} = 0.01^{+0.5}_{-0.01}$ keV.
The other best fitting  spectral parameters were
$N_{H} = (3.81^{+1.06}_{-0.97}) \times 10^{20}$ cm$^{-2}$,
$kT_{high} = 2.95^{+0.069}_{-0.026}$ keV, and
an abundance of $0.64^{+0.025}_{-0.043}$ solar, with
$\chi^2/$d.o.f.\ $ = 553.8/472 = 1.17$.
The mass-deposition rate is
$\dot{M} = 37^{+5}_{-7}~ M_{\odot}$ yr$^{-1}$, which is again
lower than the value measured with previous X-ray observatories.
The fit is very similar ($\chi^2/$d.o.f.\ $ = 553.8/473 = 1.17$;
MKCFLOW+MEKAL-abs2 in Table~\ref{tab:tot})
if the low temperature cut-off is fixed at a very low value
($kT_{\rm low} = 0.001$ keV).
The mass-deposition
rate is unchanged.
The mass of gas cooling down to very low
temperatures for the free-absorption case
is $30 < M_{\odot} < 42$ yr$^{-1}$.

We fitted the spectrum with the VMEKAL model, coupling together elements
as in the fixed-absorption case.
The best fit gave
$\chi^2/$d.o.f.\ $ = 497.2/469 = 1.06$,
$N_{H} = (0.00^{+0.76}_{-0.00}) \times 10^{20}$ cm$^{-2}$,
and a temperature of $kT = 2.86^{+0.05}_{-0.05}$.  
The Si abundance was $0.70^{+0.07}_{-0.09}$ solar, 
the S  abundance was $0.86^{+0.11}_{-0.12}$ solar, and the Fe
abundance was $0.56^{+0.05}_{-0.03}$ solar.

\subsubsection{Total Spectrum: Summary}

For all of the fits to the total spectrum, consistent values are measured
for the temperature, abundance, and mass-deposition rate.  The temperature of
the majority of the gas in the cluster is 
$kT \approx 2.9$ keV, the abundance is approximately 0.6 times the solar
value, and the mass-deposition rate found with the cooling flow models is
$26 < M_{\odot} < 42$ yr$^{-1}$.
This mass-deposition rate is approximately a factor of three lower than 
the values determined from earlier observations with {\it Einstein}, 
{\it ROSAT}, and {\it ASCA}.  
It is still much larger than the 
mass of gas we would expect to be ejected from stars in the cD galaxy.  Using
$M_B = -22.65$ (de Vaucouleurs et al. 1991) for the cD galaxy (UGC 09799), 
we find a stellar mass loss
rate of $\dot{M}_{*} = 2.7~M_{\odot}$ yr$^{-1}$ using
$\dot{M}_{*}/L_{B} = 1.5 \times 10^{-11}~M_{\odot}~\rm{yr}^{-1}~L_{\odot}^{-1}$ (Sarazin 1990).

Spectral models that include more than one temperature component 
give better fits than single-temperature models.  Except for the 
single-temperature MEKAL fit, models that allow the
absorption to vary do not give significantly better fits than the fixed 
Galactic absorption models, so there is no evidence of excess absorption in
the cluster.
However, there is still some uncertainty because of the calibration of the 
ACIS-S response at low energies.

\subsection{Spectral Profiles}

The large scale distribution of the X-ray emitting gas in Abell 2052 is
roughly circular.
We fitted the spectra extracted from seventeen circular annuli with average
radii ranging from $12\arcsec$ to $322\arcsec$, centered on (but excluding)
the central point source.
Initially, we fitted each spectrum with a single-temperature
MEKAL model with absorption fixed to the Galactic value of 
$2.85 \times 10^{20}$ cm$^{-2}$ (Dickey \& Lockman 1990).
The single-temperature model provided a 
good fit for the outer annuli, but was not adequate for the very 
inner regions of the cluster.  This is at least partly due to the hot gas from
the outer regions of the cluster being projected onto the inner regions.
Additionally, there could be a physical mixture of gas at different
temperatures found in the inner regions.

\subsubsection{Deprojection}

To better determine the physical state of the gas in the inner regions of the
cluster, we performed a spectral deprojection assuming spherical symmetry.
The spectrum from the
outermost annulus was fitted with a single-temperature MEKAL model with the 
absorption fixed to the Galactic value.  Then, the next annulus in was fitted.
The model used for this annulus was a combination of the best-fitting model 
of the exterior annulus with
the normalization scaled to account for the spherical projection of the 
exterior shell onto the inner one, along with another MEKAL component added
to account for the emission at the radius of interest.  This 
process was continued inward, fitting one spectrum at a time, so that the
model used for the innermost spectrum included seventeen MEKAL components, 
with sixteen of them fixed.
The annuli were chosen so that at least $25\%$ of the emission from each
projected annulus came from the spherical region with the same radii
(e.g., the region of interest for that annulus) with the remainder
coming from the overlying, projected emission.

\subsubsection{Temperature Profile}

\begin{figure}
\plotone{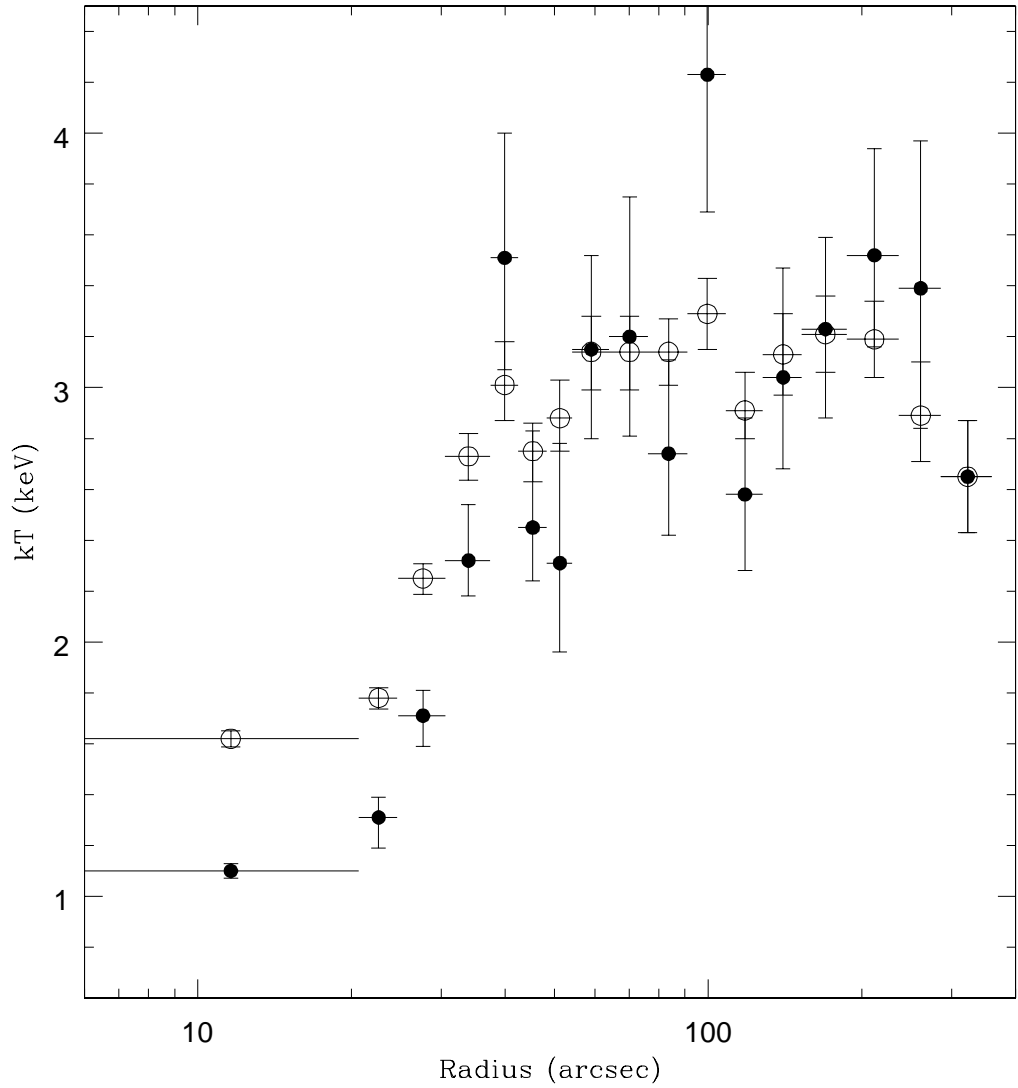}
\caption{Temperature as a function of radius in the {\it Chandra} 
observation of Abell 2052.  Temperatures obtained from single-temperature
(projected) fits are shown as open circles, and temperatures obtained from
deprojection are shown as filled circles.
The error bars are at the $90\%$ confidence level.
\label{fig:kT}}
\end{figure}

Temperature values resulting from the single-temperature fits and the 
deprojection 
are displayed in Table~\ref{tab:kT}.  The deprojection fits are better 
(lower $\chi^2$/d.o.f.) than
the single-temperature fits for only the innermost radii, out to a radius
of approximately
$30\arcsec$.  The substructure (shells and voids) seen in the X-ray image,
and associated with the radio source, is largely contained within a radius of
$30\arcsec$.
Exterior to this radius, the emission in each annulus is well-described by a 
single-temperature model and there is no evidence for multi-temperature
gas occurring there.
A plot of the projected (open circles) and deprojected (filled circles)
temperature profiles are shown in Figure~\ref{fig:kT}.
In both cases, there is a dramatic decline in the 
temperature of the intracluster medium towards the center of the cluster.
In the deprojected case, the temperature drops to a lower value at the center
than in the projected case (where the projected hotter gas raises the apparent
average temperature).
The temperature drops from an average $kT \approx 3$ keV at the outer 
annuli to $kT \approx 1$ keV at the cluster center.

\subsubsection{Abundance Profile}\label{sec:aprof}

\begin{figure}
\plotone{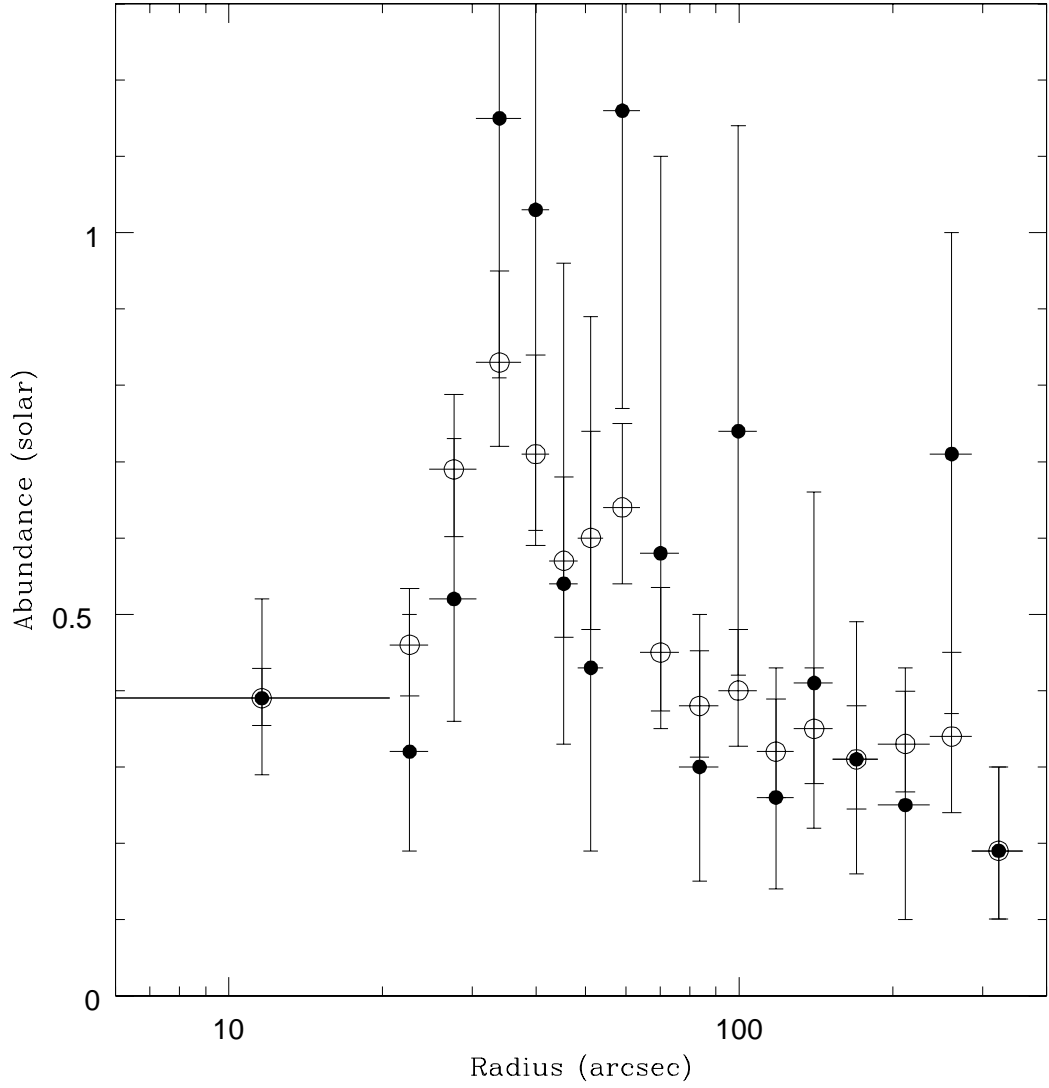}
\caption{Abundance profile for Abell 2052.
Abundance values determined using 1-temperature spectral fits are shown
with open circles, and those found
using the deprojected spectral fits are displayed as filled circles.
The error bars are at the $90\%$ confidence level.
\label{fig:abund}}
\end{figure}

The elemental abundance is approximately 0.3 times the solar value in the 
outer regions of the cluster, rises to a peak value
approximately 1.0 times solar at approximately 30 arcsec, and then drops 
again towards the center of the cluster to approximately 0.4 times solar.
This behavior is seen in both the projected (open circles) and deprojected 
(filled circles) spectral fits, as displayed in Figure~\ref{fig:abund},
and tabulated in Table~\ref{tab:kT}.
With the projected spectral fits, one could worry that since the 
single-temperature 
models for the inner few annuli do not adequately represent the data, 
the lower abundances
found for these annuli could result from XSPEC trying to compensate for the absence
of more than one temperature in the model.  Resolving this problem was part 
of our motivation for deprojecting the spectra.  After the deprojection, the
goodnesses of the fits for the inner few annuli were greatly improved, and the
overall shape of the abundance profile was consistent with that found using 
the single-temperature fits.  Therefore, this shape appears to be real, and 
not just a result of poor-fitting models in the inner regions of the cluster.

Similar abundance profiles, that rise from the outside in and then drop again
in the inner regions of the clusters, have been seen recently in 
Virgo/M87 (B\"ohringer et al.\ 2001),
Centaurus (Sanders \& Fabian 2002), Abell 2199 (Johnstone et al.\ 2002), 
and the cluster associated with 4C+55.16 (Iwasawa et al.\ 
2001).  One possible explanation for this abundance profile shape is that the 
ICM is very chemically inhomogeneous on small scales ($\sim$1 kpc), as
suggested by Fabian et al.\ (2001).
Simulations by Morris \& Fabian (2001), assuming the ICM is composed of
pure H and He regions which contain 90\% of the mass and of regions
with five times solar abundance gas containing 10\% of the mass,
were successful in reproducing an abundance
profile that rises from the outside in, with a drop in the central few tens of
kpc. 
At high temperatures, the cooling is dominated by thermal bremsstrahlung
from the H/He gas.
Below approximately 2 keV, line cooling dominates.
The metal-rich gas radiatively cools more quickly than the metal-poor gas
at all temperatures, and so reaches low temperatures before the metal-poor gas.
Line-emission cooling then dominates for this gas and there is an apparent
increase in abundance.
This happens first towards the center of a cluster where the gas is the most
dense.
The apparent central drop in abundance occurs because there is a limit to 
the energy content of the metal-rich gas, and it cools rapidly by 
line-emission to low temperatures.  
However, it should be noted that such an inhomogeneous ICM is difficult to
achieve (cf. Mathews 1990).  Also, B\"ohringer et al.\ (2002) show that the
abundance drop towards the center of the Virgo cluster cannot be 
explained by an inhomogeneous ICM.

Another possible explanation for the central drop in abundance is
resonance scattering of line emission.  This may occur in the dense cores
of clusters of galaxies (cf. Gil'fanov et al.\ 1987), where the optical 
depth of some lines, including some of the Fe L-shell lines, can become 
relatively high.
The emission is scattered to regions of lower optical depth, away from the
cluster center.
This was suggested as an explanation for the central abundance drop seen
in the Virgo cluster (B\"ohringer et al.\ 2001) using
moderate-resolution spectra from the {\it XMM-Newton} EPIC-pn and MOS. 
Examination of individual lines with high-resolution grating spectroscopy of 
the same cluster with {\it XMM-Newton} revealed, however, no evidence of
strong emission lines being redistributed by resonance scattering
(Sakelliou et al.\ 2002).  It is
possible that turbulence (possibly arising from radio jets and lobes 
interacting with the ICM) could reduce the optical depth of lines, thus
reducing any effects from resonance scattering.

\subsubsection{Absorption Profile}

We also allowed the absorption to vary in fitting the annular spectra.
Absorption ($N_H$) values for each annulus, derived from single 
temperature (1-MEKAL) fits, are given in Table~\ref{tab:nh}.
For all of the annuli except one,
the best-fitting absorption value is below the Galactic value.
This is most likely an indication of the uncertainty of the
ACIS response at low energies, where absorption has the largest effect.
The ACIS response at low energies is particularly uncertain because
of the time-dependent degradation in the quantum efficiency at low energies
and the uncertainties in the correction for this effect.
Alternatively, there might be excess soft emission coming from the cluster.
Given the uncertainty of the measurements, we hesitate to draw any conclusions
at this point from the absorption profile.  Values from a deprojection
would have even larger error bars, so we have only performed the deprojection
with absorption fixed to the Galactic value.

\subsection{Central Source}

We extracted the spectrum for the central source using an aperture 
with a radius of $2\farcs5$.
The background was determined locally from an annulus centered on 
the source with inner and outer radii of $2\farcs5$ and $5\farcs0$, 
respectively.  After background subtraction, there were 940 counts in the
source region in the 0.7 -- 7.0 keV range.

\begin{figure}
\vskip6.5truein
\includegraphics{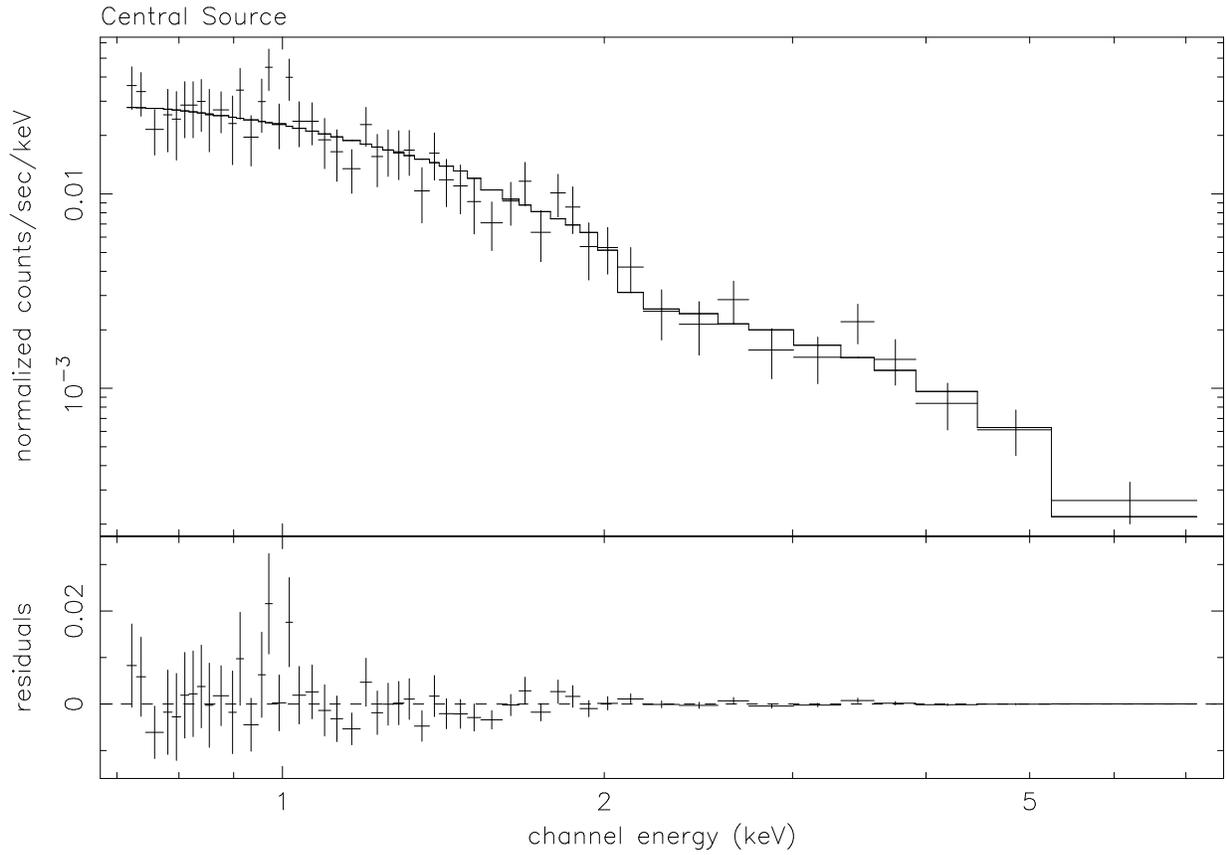}
\caption{The spectrum extracted from the central AGN, fitted with a 
model including Galactic absorption and a power-law component.  There 
is no excess
absorption above the Galactic value and the best fitting photon index is
$\Gamma = 2.00^{+0.15}_{-0.15}$.
\label{fig:AGN}}
\end{figure}

The spectrum was fitted with a model including absorption and a power-law.
Freeing the absorption did not improve the fit, so it was fixed to the 
Galactic value.  The best-fitting photon index was 
$\Gamma = 2.00^{+0.15}_{-0.15}$, which is typical for a radio galaxy
(e.g., Sambruna, Eracleous, \& Mushotzky 1999).
The spectrum along with the best-fitting model is shown in
Figure~\ref{fig:AGN}.
The unabsorbed flux in the 0.7 -- 7.0 keV range is
$1.5 \times 10^{-13}$ erg cm$^{-2}$ s$^{-1}$, and the X-ray luminosity in
this same energy band is $L_{X} = 7.9 \times 10^{41}$ erg s$^{-1}$.

In order to constrain the amount of internal absorption, 
we also fitted the spectrum with a model including Galactic absorption,
a power-law, and an extra internal absorption (``ZWABS'') component.
The best fitting
value for any internal absorption was $0.0^{+4.44}_{-0.0} \times 10^{20}$
cm$^{-2}$.
Thus, there is no evidence for extra absorption associated with the AGN,
with the upper limit being $4.4 \times 10^{20}$ cm$^{-2}$.

\section{Mass Profile}

\begin{figure}
\plotone{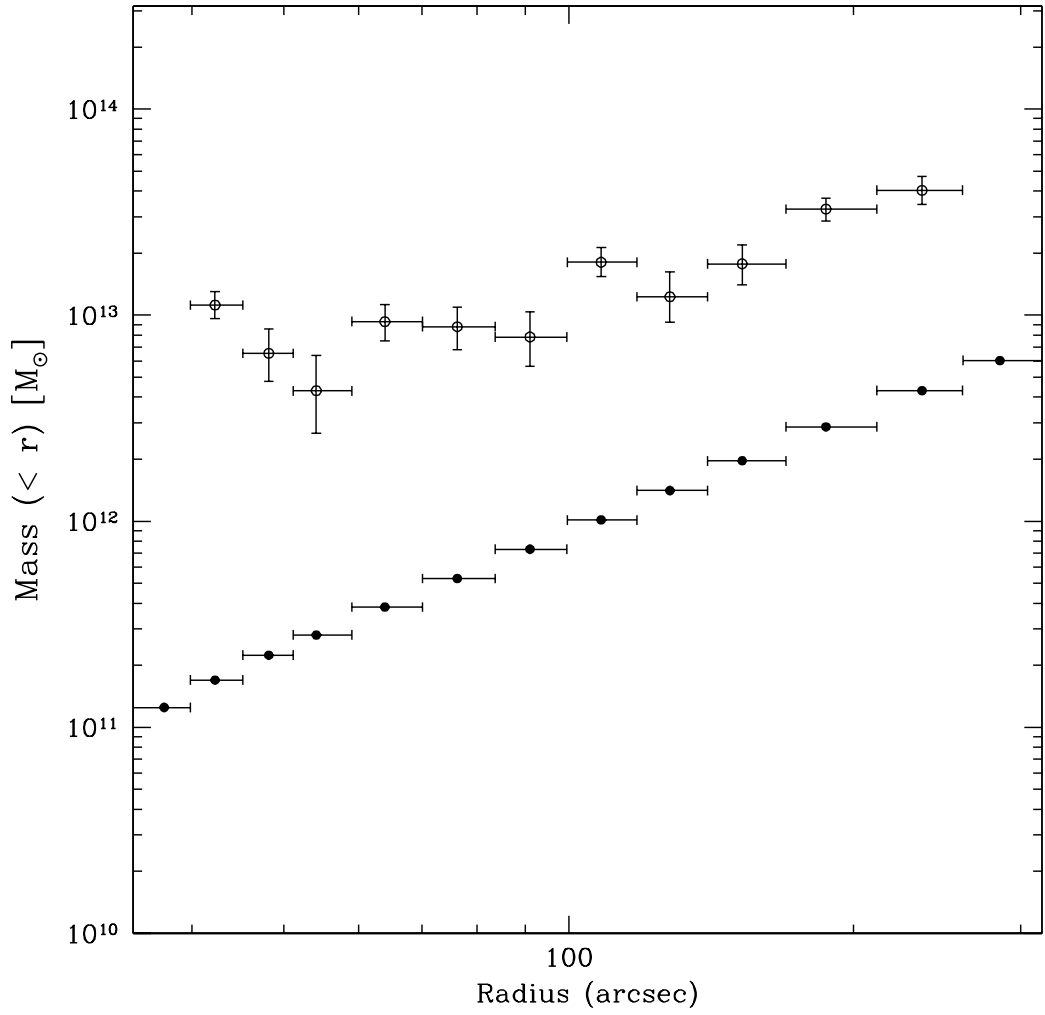}
\caption{The total mass (open circles) and gas mass (filled circles)
profiles for Abell 2052.
The error bars are at the $1-\sigma$ level.
\label{fig:mass}}
\end{figure}

The gas mass (filled circles) and total mass (open circles) profiles for
Abell 2052 were determined from the X-ray observation and are displayed
in Figure~\ref{fig:mass}.
The X-ray surface brightness profile was deprojected to give the
emissivity and density as a function of radius, assuming spherical symmetry.
The gas mass is then given by
\begin{equation} \label{eq:mgas}
M_{\rm{gas}}(<r) = \int_{0}^{r} 4\pi r^{2} dr \rho_{\rm{gas}}(r) \, .
\end{equation}

The total mass is given by the equation of hydrostatic equilibrium:
\begin{equation} \label{eq:mtot}
M_{\rm{tot}}(<r) = -\frac{kTr}{\mu m_{p}G}
\left( \frac{d~\ln~\rho_{\rm{gas}}}{d~\ln~r} 
+ \frac{d~\ln~T}{d~\ln~r} \right) \, .
\end{equation}
In applying the hydrostatic condition, we consider only the radii
$>$40\arcsec;
at smaller radii, the radio bubble structure indicates that the gas
is not hydrostatic.
We take the temperature ($T$) values
from our single-temperature (1-MEKAL) spectral fits.  These fits are as
good as multi-temperature (deprojected) fits for all but the innermost 
annuli (which are not used in the mass profile), and the temperatures have 
less scatter and smaller errors than those found using the deprojected fits.
Errors for both the gas and total mass were calculated using propagation of
errors.
The gas mass fraction increases from approximately 5\% in the inner
regions to 10\% in the outer radii ($\approx 250$ kpc);
this is similar to results found on similar scales in other clusters
(David, Jones, \& Forman 1995; Allen, Schmidt, \& Fabian 2002).

\section{The Radio Source Interaction Region}

\begin{figure}
\plotone{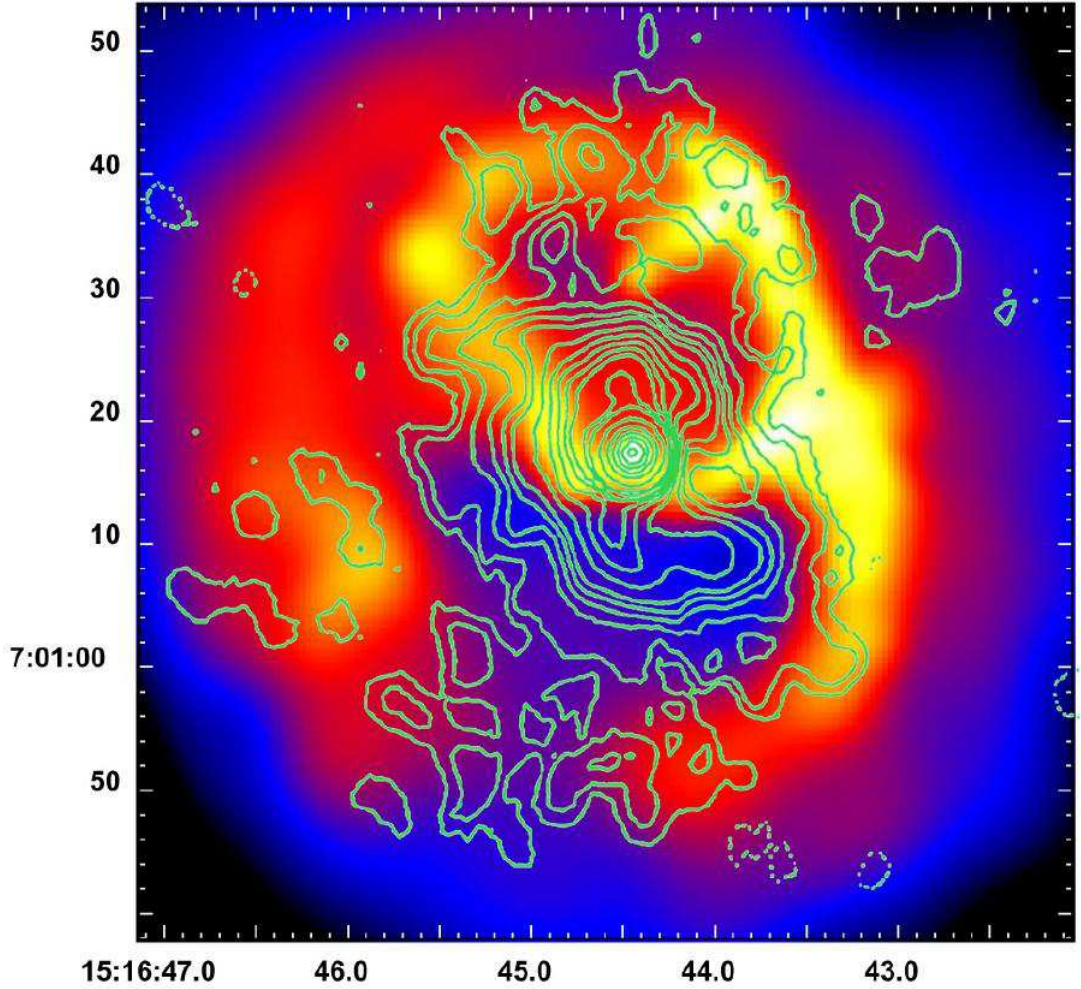}
\caption{Adaptively smoothed Chandra ACIS-S3 image of the central region
of Abell 2052 with radio contours (Burns 1990) superposed.  The radio 
source has swept out ``holes'' or ``bubbles'' in the X-ray emitting gas,
creating bright shells of compressed X-ray gas surrounding the holes.}
\label{fig:radio}
\end{figure}

We now discuss several topics concerning the center of the cluster,
where the X-ray emitting gas is greatly affected by the central radio
source 3C 317.
An adaptively smoothed image of the central region is displayed in 
Figure~\ref{fig:radio}, with radio contours (Burns 1990) superposed.  The
radio source has swept material out of the center of the cluster creating
two ``holes'' or ``bubbles'' in the X-ray emission, and this material has
been compressed into bright shells surrounding the holes.  There is a
``bar'' of bright material passing east-west through the cluster center that
is likely the intersection of the northern and southern shells.  There
is also a ``spur'' of emission that protrudes into the northern bubble.

\subsection{Spatial Distribution of Temperature and Abundance}

In order to explore the smaller scale distributions of temperature and 
abundance
at the cluster center, and particularly, how they relate to the structure 
of the bright X-ray shells, we created maps of these quantities for the 
central $74 \times 74$ arcsec region of the cluster.  Each map includes
a grid of $25 \times 25$ boxes, where each box has dimensions of 
$6 \times 6$ pixels 
(approximately $3 \times 3$ arcsec).  A spectrum was extracted from each box,
with the requirement that it include at least 900 counts for a spectral
fit to be performed.  Background was taken from the blank sky fields of
Markevitch\footnotemark[4].  Each spectrum was then fitted in the
0.7--8.0 keV range with a model
combining absorption fixed to the Galactic value and a single-temperature
MEKAL model.

\begin{figure}
\plotone{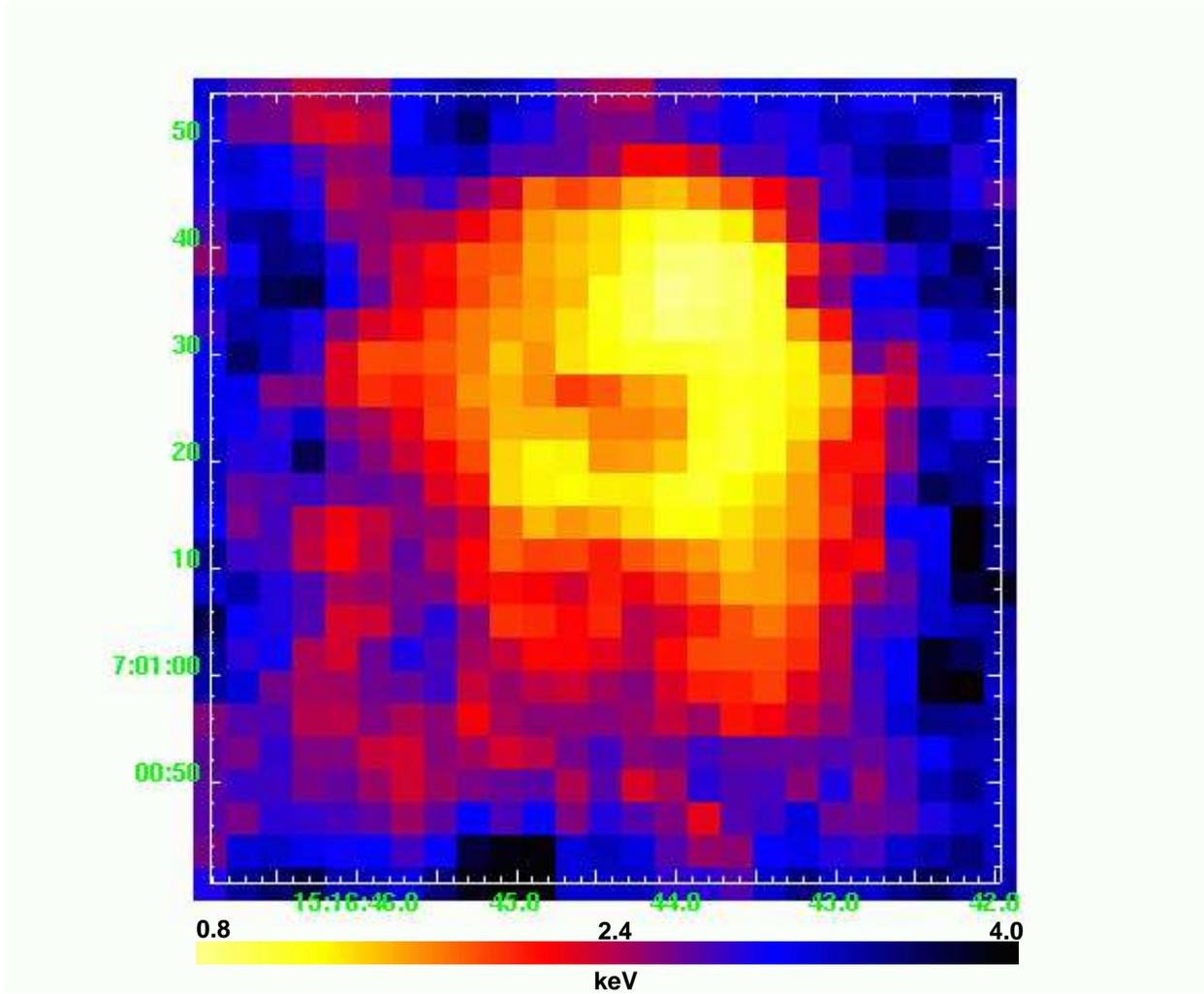}
\caption{Temperature map of the central region of Abell 2052.  The coolest
regions (in projection) correspond with the brightest parts of the X-ray
shells that surround the radio source.}
\label{fig:tmap}
\end{figure}

The temperature map is shown in Figure~\ref{fig:tmap}.  This map shows that
the coolest gas is found in the center of the cluster.  The gas surrounding 
the radio source in the shells is cool, rather than hot, as would 
be expected if it was strongly shocked (Heinz, Reynolds, \& Begelman 1998).  
Gas as cool as 0.8 keV
is found coincident with the brightest parts of the X-ray shells -- the bar
that runs east-west through the cluster center, the western and northwestern
portions of the northern shell, and the spur of emission that protrudes
into the northern shell.  These are the same regions that are emitting in
H$\alpha$ (Paper I).  It is possible that the very coolest gas is found in
these regions of the shells. It is also possible, since these regions are 
also the brightest X-ray emitters, and the maps indicate the temperature of 
the gas along the line-of-sight through the cluster (including projected,
hotter, gas), that these parts of the shells are measured to be coolest 
because they are the regions of the shells where the largest fraction of
projected emission is coming from the shells rather than overlying emission.

\begin{figure}
\plotone{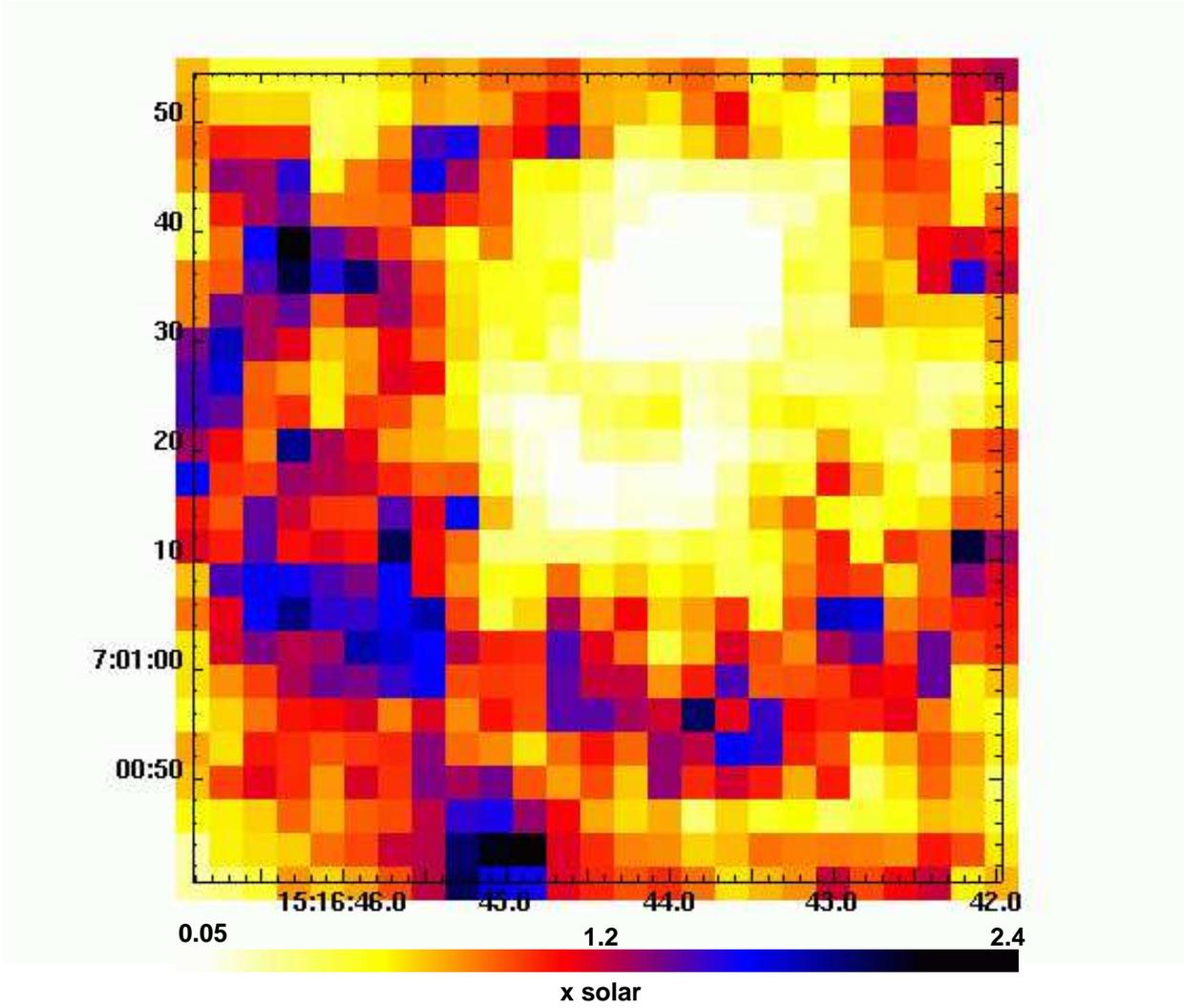}
\caption{Abundance map of the central region of Abell 2052.  The least 
abundant regions of the cluster are coincident with the brightest part of
the X-ray shells.  The abundance rises exterior to the bright, X-ray shells.
The lowest abundance regions are white.}
\label{fig:amap}
\end{figure}

The abundance map is displayed in Figure~\ref{fig:amap}.  This map
confirms what we saw in the abundance profile in \S\ref{sec:aprof}:  
after rising
from the outside of the cluster inward, the elemental abundance drops again
at the very center of the cluster.  The peak seen in the abundance profile
occurs exterior to, and not coincident with, the X-ray bright shells.  In
fact, the lowest abundances are measured in
the same regions of the cluster that the lowest temperatures are measured,
namely, the brightest regions of the shells.
This is intriguing, because it may support the idea that the gas may be
chemically inhomogeneous on small scales 
(Fabian et al.\ 2001 ; Morris \& Fabian 2001). 
In the low temperature regions,
the high-metallicity gas has already cooled below X-ray emitting
temperatures, leaving behind only the low-metallicity gas.  We are cautious
in this interpretation, however, because temperature and abundance are
correlated when doing spectral fitting in the sense that either a low
temperature or a high abundance can account for strong emission at low 
energies.

\subsection{Shell Masses} \label{sec:shmass}

Here, we update the value given in Paper I for the mass of gas in the 
bright shells of emission surrounding the X-ray holes.  The new values
do not change any conclusions in Paper I, but are more accurate.
The masses are still
consistent with the shells being formed by gas compressed out of the bubbles
by the radio source, and with the bubbles being devoid of X-ray emitting gas.

We approximate the southern shell as a sphere, centered on the southern 
bubble, with inner and outer radii
of $11\farcs9$ (11.3 kpc) and $21\farcs9$ (20.8 kpc), respectively.  We use
a density of $n_e = 0.035$ cm$^{-3}$ which is the average of the density value
calculated from the surface brightness of the very bright portion of the 
ring to the west of the center of the cluster ($n_e = 0.04$ cm$^{-3}$),
and the density value taken from the density profile (Paper I)
at the shell radius ($n_e = 0.03$ cm$^{-3}$).  In Paper I, we
used the density determined from the bright portion of the ring which was
probably an overestimate since that is the brightest, and therefore most
dense, portion of the southern shell.  Also, in Paper I, we estimated the
size of the shell to be somewhat larger.  For the updated measurement, we
calculate a mass of $3.1 \times 10^{10} M_{\odot}$ for the southern shell.
The predicted mass for this shell, if all of the mass came from evacuating
material out of the southern bubble by the radio source, and extrapolating
the density profile to the center of the bubble, is
$(5.4\pm2.7) \times 10^{10} M_{\odot}$.
Therefore, the mass of the shell is
consistent with the southern X-ray ``hole'' or ``bubble'' 
being devoid of X-ray gas, with all of it being swept up in the shell.

We performed the same analysis for the slightly smaller northern shell.
In this case, the inner and outer radii are $7\farcs9$ (7.5 kpc) and
$17\farcs9$ (17.0 kpc), respectively.  The mass for the northern shell is
$1.9 \times 10^{10} M_{\odot}$, with a predicted mass of
$(1.6\pm0.7) \times 10^{10} M_{\odot}$.  Again, this is consistent with the
northern shell being formed from material pushed out of the northern bubble
by the radio source and compressed in the shell, and with the bubble being
devoid of X-ray emitting gas.
The total mass for the northern and southern shells combined is 
$\approx 5.0 \times 10^{10} \, M_{\odot}$.

\subsection{Limiting the Temperature of Thermal Gas in the Bubbles}

As discussed in Paper I, the pressure in the X-ray shells
($P = 1.5 \times 10^{-10}$ dyn cm$^{-2}$) is
considerably higher
than the pressure derived in the X-ray holes from the radio observations, 
assuming equipartition
of energy
($P = [2 - 5] \times 10^{-11}$ dyn cm$^{-2}$; Zhao et al.\ 1993).  One
possible resolution of this discrepancy is that the X-ray holes are
filled with very hot, diffuse, thermal gas which provides the necessary
additional pressure.

We attempted to limit the temperature of any thermal component which
supplies the bulk of the pressure in the X-ray holes.
We first tried to derive such a limit from the
${\it Chandra}$ X-ray spectrum extracted from the southern X-ray hole.
First, we fitted the spectrum with a two-temperature MEKAL model.  We then
added an additional MEKAL component to the fit, with the normalization
(which is a function of density) fixed so that the pressure of the 
additional component was $P = 1.5 \times 10^{-10}$ dyn cm$^{-2}$.
This was done for gas at temperatures of 20, 10, and 5 keV.  No significant
change in the goodness of the fit, as judged by a change in the
$\chi^{2}$ value, was observed for any of the temperatures.  We therefore
cannot place any limits, using this method, on the temperature of a hot,
diffuse, thermal gas component that may be present in the holes and which
might provide their pressure support.

Using another technique, we find a more restrictive limit on the
temperature of hot thermal gas which might provide the bulk of the pressure
support in the X-ray holes.
We use the radio observations of Zhao et al.\ (1993) and Ge \& Owen (1994), and
the lack of Faraday depolarization observed in the X-ray holes to place 
this limit.
Zhao et al.\ (1993) define several regions of the radio source, including
those they call ``bipolar'' (a region 30$^{\prime\prime}$ tall by 
15$^{\prime\prime}$ wide centered
on the AGN) and ``halo'' (a region 75$^{\prime\prime}$ tall by 
45$^{\prime\prime}$ wide centered on
the AGN).  The X-ray holes cover a region intermediate in size between these
two.  The equipartition radio pressure calculated for the 
halo component is $P_{\rm {eq}} = 2 \times 10^{-11}$ dyn cm$^{-2}$, and that 
for the
bipolar region is $P_{\rm {eq}} = 5 \times 10^{-11}$ dyn cm$^{-2}$.  We 
therefore adopt $P_{\rm {eq}} \approx 3 \times 10^{-11}$ dyn cm$^{-2}$ for the
X-ray holes.
Zhao et al.\ (1993) also calculate equipartition magnetic fields of
$B_{\rm {eq}} = 10~\mu$G (halo region) and 
$B_{\rm {eq}} = 20~\mu$G (bipolar region).
We adopt $B_{\rm {eq}} = 15~\mu$G for the X-ray holes.
Large Faraday rotation measures (RMs; Ge \& Owen 1994) were determined, and
average polarizations for the north and south X-ray holes of $\approx 30\%$ 
(at 3.6 cm) and $\approx 15\%$ (at 6 cm) were measured.
In order to produce large RMs without depolarization, the magnetized
plasma must lie in front of, and not within, the radio emission regions.
We can therefore place a limit on the rotation measure due to plasma within
the radio emission region:
\begin{equation} \label{eq:phi}
\phi = RM \, \lambda^{2} \la \frac{\pi}{2} \, ,
\end{equation}
where $\phi$ is the angle of rotation, $RM$ is the rotation measure in
rad m$^{-2}$, and $\lambda$ is the wavelength of the radiation in meters.
At 6 cm, then, the rotation measure is $RM \le 440$ rad m$^{-2}$. The
rotation measure is defined as:
\begin{equation} \label{eq:rm}
RM = 8.12 \times 10^{5} \int n_{e} B_{\Vert} dl
\, {\rm rad} \, {\rm m}^{-2} \, ,
\end{equation}
where $n_{e}$ is the electron density in cm$^{-3}$, $B_{\Vert}$ is the magnetic
field along the line of sight in Gauss, and $l$ is the path length along the
line of sight in parsecs.  This can also be written:
\begin{equation} \label{eq:rm2}
RM = 244
\left(\frac{n_{e}}{10^{-3}~\rm{cm}^{-3}}\right)
\left(\frac{B_{\Vert}}{15~\mu \rm{G}}\right)
\left(\frac{l}{20~\rm{kpc}}\right)
\, {\rm rad} \, {\rm m}^{-2} \, .
\end{equation}
The density is then limited to:
\begin{equation} \label{eq:dens}
\left(\frac{n_{e}}{10^{-3}~\rm{cm}^{-3}}\right) \la 1.81 \left(\frac{B_{\Vert}}{15~\mu \rm{G}}\right)^{-1} \left(\frac{l}{20~\rm{kpc}}\right)^{-1} \, .
\end{equation}
Note that this limit applies to any thermal gas in the X-ray holes, whether
it supplies the bulk of the pressure or not.
The thermal pressure is limited to:
\begin{equation} \label{eq:press}
P_{\rm {therm}} \la 5.70 \times 10^{-11}
\left(\frac{B_{\Vert}}{15~\mu \rm{G}}\right)^{-1}
\left(\frac{l}{20~\rm{kpc}}\right)^{-1}
\left(\frac{T}{10~\rm{keV}}\right)
\, {\rm dyn} \, {\rm cm}^{-2} \, .
\end{equation}
The extra pressure necessary to support the X-ray shells is the 
difference between the pressure measured in the X-ray for the shells 
and the equipartition 
pressure measured in the bubble region using the radio observations:
\begin{equation} \label{eq:press2}
P_{\rm {therm}} = P_{\rm {X\-ray}} - P_{\rm {eq,radio}}
\ge 1.2 \times 10^{-10} \, {\rm dyn} \, {\rm cm}^{-2} \, .
\end{equation}
Using equation (\ref{eq:press}), a bubble diameter of 20 kpc, 
and a $B_{\Vert}$ value of 15 $\mu$G, the 
temperature of thermal gas in the bubbles (holes) is $kT \ga 20$ keV, assuming
it provides the pressure support.

\subsection{Magnetic Field and Pressure in the X-ray Gas}

The large Faraday rotation measures in 3C~317 can be used to estimate
the magnetic field strengths and pressures.
As noted above, the magnetized plasma must not lie within the radio emission
regions, but in front of them.
We consider two locations for the magnetoactive gas:
the front edges of the compressed shells of cooler gas surrounding the
radio holes, or the undisturbed cooling flow gas at larger radii.
Ge \& Owen (1994) found RM values ranging from $-1000$ to 1000 rad m$^{-2}$
which varied on scales of $l_B \approx 3$ kpc.
Assuming our projected density for the X-ray shells of $n_e \approx 0.035$
cm$^{-3}$ and a thickness
along the line of sight of $l \approx 10$ kpc (consistent with the widths of the
edges of the shells), the required average of the magnetic field along
the line of sight in the shells is then
$| \langle B_{\Vert} \rangle |  \approx 3.5$ $\mu$G
using equation (\ref{eq:rm}).
If the magnetic field were aligned along the line of sight, this would
imply a magnetic pressure of only
$P_{B} \approx 5 \times 10^{-13}$ dyn cm$^{-2}$,
which is about 300 times lower than the gas pressure.
However, the variations in the RM suggest that the magnetic field is
not uniform and aligned, and that the average value along the line of
sight underestimates the total field strength.
Assuming random variations along the line of sight on a scale $l_B \ll l$,
we expect that the average line-of-sight field is given by
\begin{equation} \label{eq:Bpara}
\langle B_{\Vert} \rangle^2 \approx \frac{1}{3} \,
\left( \frac{l_B}{l} \right) \langle B^2 \rangle \, ,
\end{equation}
where the factor of 1/3 accounts for the three component of the field.
For the magnetic field in the shell, this gives an r.m.s.\ value
of $\langle B^2 \rangle^{1/2} \approx 11$ $\mu$G, and a pressure
of $P_{B} \approx 5 \times 10^{-12}$ dyn cm$^{-2}$.
This is still 30 times smaller than the thermal pressure.

Alternatively, we consider the possibility that the Faraday rotation is
due to gas in the external undisturbed cooling flow region.
We adopt the deprojected electron density and pressure distributions
determined from the X-ray surface brightness (Paper I).
We will assume that the magnetic field varies such that the ratio of
magnetic pressure to gas pressure is constant, although the results are
not strongly dependent on this.
Integrating the rotation measure from the deprojected gas distribution,
we find that the average value of the line-of-sight component of the field
is
$| \langle B_{\Vert} \rangle |  \approx 1.9$ $\mu$G
just outside of the X-ray shells.
If we assume that the field is tangled on a constant scale of
$l_B \approx 3$ kpc,
this implies that the r.m.s.\ magnetic field is
$\langle B^2 \rangle^{1/2} \approx 16$ $\mu$G.
The associated pressure is $P_{B} \approx 1 \times 10^{-11}$ dyn
cm$^{-2}$, which is about
15 times smaller than the gas pressure in
the cooling flow region immediately outside the shells.

Since the Faraday rotation presumably occurs primarily in only one of
these two locations, these values can be viewed as upper limits on the
field and magnetic pressure in each region.
These values indicate that, while strong magnetic fields exist in
the central regions of the cooling flow in Abell~2052, they are still
significantly weaker than equipartition.

\subsection{The Nuclear Region}

\begin{figure}
\plotone{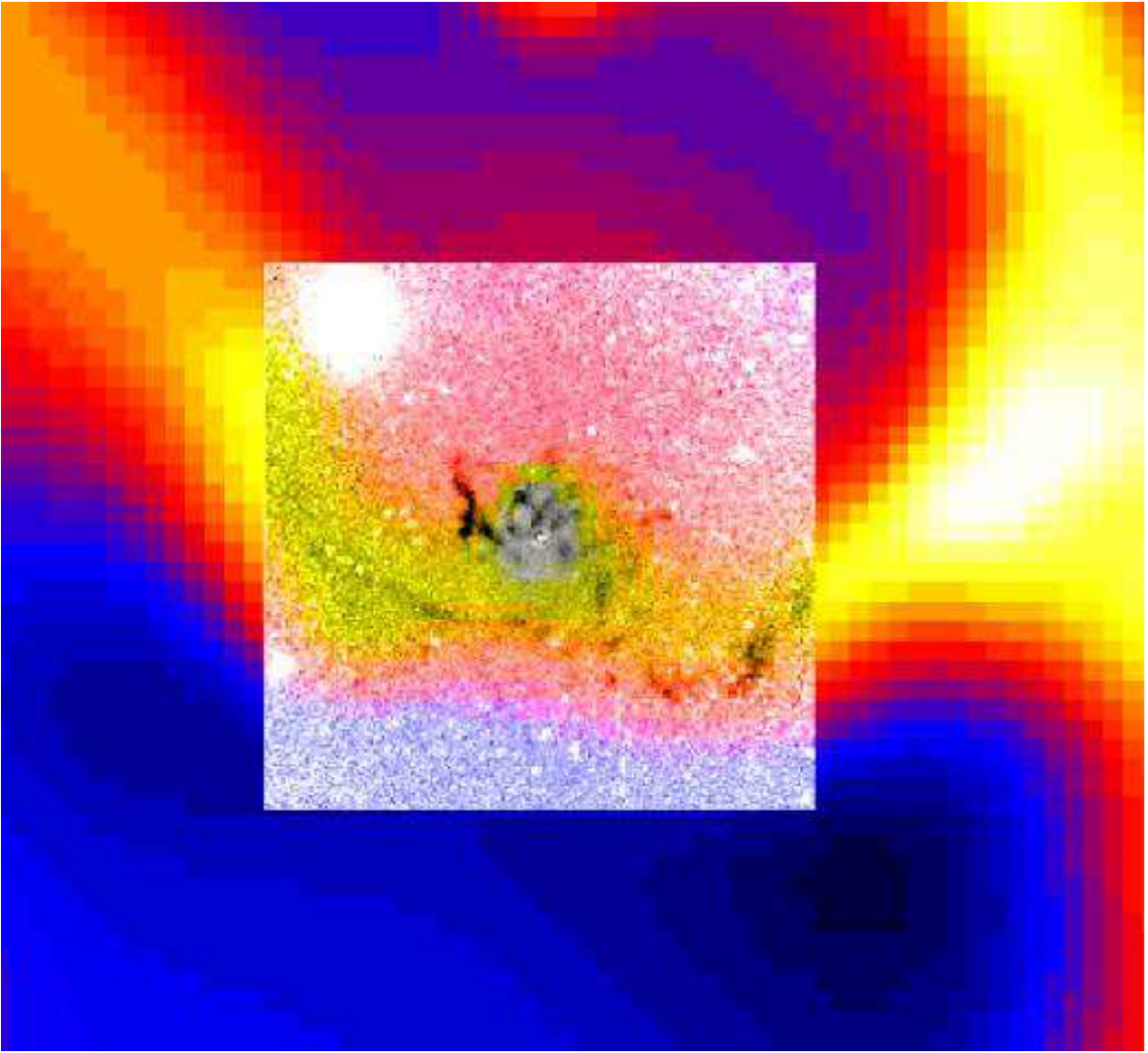}
\caption{Overlay of dust (gray/black in small box) seen with HST (Sparks 
et al.\ 2000) onto the 
adaptively smoothed X-ray emission (color) of the center of Abell 2052.  
The central, small box containing the HST observation is $13\farcs8 \times 
13\farcs8$.
\label{fig:dust}}
\end{figure}

At the center of the cluster, there is a
bar of X-ray and optical line emission (Paper I; Baum et al.\ 1988)
running across the nucleus of the AGN.
This may simply be part of the northern X-ray shell, and may only be
near the nucleus in projection.
In support of this, we note that the nucleus appears to lie slightly above
the brightest part of the bar, and that the X-ray spectrum of the
nucleus doesn't show any excess absorption.
Alternatively, the bar may be part of a disk of material at the center of
the cD galaxy, which eventually feeds the accretion disk in the AGN.
A similar disk is seen in Hydra A
(McNamara et al.\ 2000).
In support of this idea, we note that the optical emission line spectra
along the bar show evidence for velocity shear of $\sim$$\pm$80 km s$^{-1}$
about the nuclear velocity
(Heckman et al.\ 1989),
as might be expected for a rotating disk.
Also, dust was detected in {\it Hubble Space Telescope} images
of the central regions
(Sparks et al.\ 2000);
this dust appears to surround the AGN and lie at the edges of the X-ray bar
(Figure~\ref{fig:dust}).

\subsection{Northern X-ray/H$\alpha$ Spur}

The optical emission line image of Abell~2052 
(Baum et al.\ 1988; see also Paper I, Fig.~4)
shows a feature $\sim$15\arcsec\ north of the AGN.
This feature is coincident with a bright spur of X-ray emission in the
{\it Chandra} image.
Based on the H$\alpha$ emission, Baum et al.\ had suggested that this
feature might be connected with the region of line emission curving
below the nucleus of the central cD, and that this linear feature might be
due to tidal stripping from a cluster galaxy.
However, subsequent velocity measurements
(Heckman et al.\ 1989)
showed that the H$\alpha$ spur to the north and the emission around the
AGN were kinematically distinct.
In our {\it Chandra} image, the H$\alpha$ spur appears to be the brightest
portion of the northern X-ray shell, and the optical line emission may
simply be due to cooling of X-ray gas in this dense shell.
Based on the combination of the optical and X-ray images and the optical
spectra, it seems unlikely that the northern spur is connected with the
region of line emission near the nucleus.

The optical emission lines show a continuous velocity shear across
this spur, ranging from about $-130$ to $+130$ km s$^{-1}$, with
zero velocity being the nuclear velocity and the negative velocities
occurring on the part of the spur nearest to the AGN.
We consider two possible models for the origin of this feature.
First, it might be a high density region on the surface of the northern
X-ray shell and projected within the shell.
The sense of the velocity shear implies that the spur is located on
the front (nearer) side of the shell.
Second, it could be a radial filament produced a Rayleigh-Taylor
instability of the shell
(e.g., Soker, Blanton, \& Sarazin 2002), in this case occurring on the
far side of the shell.
In either case, the fact that the velocities are symmetric about zero
is somewhat surprising, although this is easier to understand if the
spur is on the surface of the X-ray shell.

If the shear across the spur is representative of the expansion
of the shell, we can estimate the shell expansion velocity $v_{\rm exp}$.
Let $\Delta v \approx 260$ km s$^{-1}$ be the total shear in the radial
velocity across the spur.
Let us assume that the spur is located on the surface of the northern
X-ray shell, and that it is expanding with the shell.
The length of the spur is $l_{\rm sp} \approx 14\arcsec$.
The spur is not exactly radial; the projected radial extent of the spur
is about $l_\perp \approx 12\arcsec$.
Let us assume that the northern edge of the spur coincides with the
projected outer edge of the X-ray shell.
The radius of the northern X-ray shell is $R \approx 18\arcsec$.
We assume that the shell is expanding spherically.
Then, the radial velocity shear across the spur $\Delta v_r$ and the 
expansion velocity of the shell $v_{\rm exp}$ are related by
$v_{\rm exp} \approx \Delta v_r ( 2 r l_\perp - l_{\rm sp}^2 )^{1/2} / R$,
which
leads to $v_{\rm exp} \approx 300$ km s$^{-1}$.
If the spur is not located on the outer surface of the shell or the
geometry is more complicated, we generally find that the required value of
$v_{\rm exp}$ is increased.
Calculations with a variety of geometries
suggest that the expansion velocity is
$v_{\rm exp} \sim$  300--600 km s$^{-1}$.
Based on the lack of strong shocks surrounding the X-ray shells,
we determined an upper limit on the Mach number of expansion of
$\le$1.2 (Paper I), which corresponds to a velocity limit of
$\la$900 km s$^{-1}$.

\subsection{Optical/UV Correlations with the X-ray}

The central galaxy in Abell 2052 has also been observed with $HST$ in
the wide-band optical and UV regions of the spectrum.
Features in the $HST$ R-band optical emission contours,
including two resolved sources possibly attributable to small galaxies
falling into the cD (Baum et al.\ 1988; Zirbel \& Baum 1998) 
do not correspond with any features seen in the X-ray.
A blue, nearly linear filament, seen in the $HST$ near-UV image of the cD
(Martel et al.\ 2002) does not directly correspond with any features seen in 
the X-ray either.
However, this feature is so small and narrow that any associated X-ray
feature might not be visible against the bright ambient X-ray emission
at {\it Chandra}'s resolution.
The filament is oriented north-south, with the northern portion
falling in the east-west running portion of the X-ray bright shells, to the
southwest of the nucleus.

\subsection{Overall Geometry of Central Region of Abell~2052}
\label{sec:geometry}

The presence of two well-defined X-ray holes in Abell~2052 suggests
that we are viewing the system roughly perpendicular to the axis of
the radio jets which inflated the radio bubbles.
The two radio holes are only separated from one another by the narrow bar
of X-ray emission which crosses just below the AGN.
This suggests that the two radio bubbles have expanded until they collided
at the center, and that the radii of the bubbles are nearly equal to the
distances from the AGN to their centers.
The agreement of the masses of the X-ray shells with that expected from
the interior gas and the strong brightness decrement in the centers of the
bubbles all argue that the bubbles are located at approximately their
projected distance from the AGN.
On the other hand, there are a number of arguments which suggest that we
are viewing the system at an intermediate angle to the radio axis.
The central AGN is located slightly above the central bar of X-ray
emission.
Of course, this might just indicate that the northern bubble expanded
beyond the AGN before colliding with the southern bubble.
Alternatively, the velocity shear in the H$\alpha$ associated with the
central bar may indicate that this is an accretion disk around the AGN.
The lack of excess absorption towards the AGN suggests that we
are not observing the system nearly perpendicular to the radio axis.
Also, there is no evidence for a Doppler-boosted jet (or any jets, for
that matter) in the radio images
(Zhao et al.\ 1993).
Taken together, these arguments may be most consistent with the radio
axis being at an intermediate angle $\sim$45$^\circ$ to our line of sight,
and with the radio bubble centers being at a slightly larger distance
from the AGN ($[ \sin 45^{\circ}]^{-1}\sim$1.4 times) than their radii.

\section{Star Formation}

Understanding the possible relationship between the cooling
ICM and star formation  is one of the goals of this study.
In this context, Abell 2052 is one of the best objects
with which to address this issue, as several investigations
have shown that the inner 3 kpc of the central cluster
galaxy is anomalously blue (McNamara \& O'Connell 1989, 1992;
Crawford et al.\ 1999).  This blue region
has recently been resolved by the {\it Hubble} Space Telescope 
into an extended component that is almost
certainly star formation, and an unresolved nucleus that
may be either a compact star formation region or nonthermal
AGN emission (Martel et al.\ 2002).  In addition, the 
entire blue region is enshrouded in dust (Martel et al.\ 2002).  

In order to estimate the star formation rate, or more precisely,
the luminosity mass of the young stellar population, we have 
reanalyzed the U-band imaging and spectroscopy of 
McNamara \& O'Connell (1989, 1992) combined with the recent 
{\it Hubble} results of Martel et al.\ (2002).
We estimated the fraction of the U-band light
emerging from the blue population in the inner 3 arcsec radius ($\simeq 3$ kpc)
by modeling the U-band light of the host galaxy 
with an $R^{1/4}$-law surface brightness profile with a
softened core.  The fraction of light, $f_{\rm AP}$, 
contributed by the blue ``accretion population'' was found by comparing
the model U-band image of the host galaxy alone to the real image,
which includes the host galaxy plus the accretion population.  
Our analysis shows that the excess U-band light from the
accretion population contributes an average of $\simeq 14\%$ of
the light within a 3 arcsec radius of the nucleus.
We then determined the mass of the accretion population as 
$M_{\rm AP}=M/L(U)_{\rm AP}f_{\rm AP}L(U)$, where
$M/L(U)_{\rm AP}$ is the U-band mass-to-light ratio of the accretion
population, and $L(U)$ is
the total U-band luminosity of the central blue region.
We found the U-band luminosity of the accretion population alone
to be  $L(U)_{\rm AP}\equiv f_{\rm AP}L(U)=6\times 10^8~{\rm L_{\odot}}$ 
before correcting for 
extinction, and $10^9~{\rm L_{\odot}}$ after correcting for
extinction, as  discussed below.
These population mass estimates assume that all of the excess
blue light emerging from the core of the galaxy comes from
an accretion population.  However, a small fraction of the light may
be coming from the active nucleus. In this case, our population masses
should be taken as upper bounds.

One of the most uncertain elements of the mass estimate is the 
determination of
$M/L(U)_{\rm AP}$, which is related to the star formation 
history (burst, continuous) and the age of the population.  
(In our calculations, we assume the Salpeter initial mass function with 
near solar abundances for the model stellar populations of
Bruzual \& Charlot [1993].)  This estimate is particularly
difficult for an object like Abell 2052 which has a relatively
modest central color excess of $\delta(U-B)\simeq -0.17$ 
(McNamara \& O'Connell 1989, 1992).  The intrinsically bluer
colors of the accretion population that would produce a 
color excess of this magnitude in the composite background plus
accretion population can be attributed to 
either a lower mass, young burst population, or an older, more massive 
population.   The population colors derived from 
{\it Hubble} images of Abell 2052 (Martel et al.\ 2002) provide the strongest 
constraints available on the star formation history, and these
constraints are consistent with as many as three 
young and intermediate age populations.

The youngest population is found in a
blue, low mass filament with an age of only a few Myr (Martel et al.\ 2002).
However, this $\sim 3\times 10^4~{\rm M}_\odot$ filament is only a minor
component  of the accretion population seen both from the ground and in the
{\it Hubble} images.  The blue population as a whole has
colors consistent with an aging burst population that
was deposited between $0.1-1$ Gyr ago, or constant star formation
over a period of $\sim 1$ Gyr.  The broad range in these estimates
may reflect the existence of a composite blue population
with a range of star formation histories and
ages.  Furthermore, the uncertainty
in the population history or its composite nature 
translates into an order of magnitude
range in the population's estimated mass because of the rapidly rising
U-band mass-to-light ratios for populations older than $\sim 10$ Myr
(Bruzual \& Charlot 1993).  

With these broad considerations in mind, the young population is consistent
with being an aging burst with a  luminosity mass 
$M_{\rm AP}\sim 9\times 10^7~ {\rm M}_\odot - 1.2\times 10^9~ {\rm M}_\odot$,
that occurred between  $0.1-1$ Gyr ago.  At the same time,
the colors are consistent with constant star formation
that began $\sim 1$ Gyr ago, with a star formation rate of
$\dot M_* \simeq 0.36 ~{\rm M}_\odot ~{\rm yr}^{-1}$.  Note that these
luminosity mass estimates assume no diminution or reddening of the U-band
light by the dust features seen in the {\it Hubble} 
images (Martel et al.\ 2002).  
The internal reddening by this dust was found by Crawford et al.\ (1999) to be 
$E(B-V)=0.15\pm 0.05$, based on anomalous Balmer emission line ratios
toward the dusty region.  Assuming the dust associated with the accretion
population is distributed in a foreground screen,
the color excess corresponds to a
U-band extinction of ${\rm A}(U)=0.73$ mag (see Cardelli, Clayton,
\& Mathis 1989).  Uncorrected extinction at this level would have the
effect of causing an underestimate of the luminosity masses and
star formation rates by almost a factor of 2.  After correcting for extinction,
the continuous star formation rate rises to
$\dot M_* \simeq 0.6~ {\rm M}_\odot ~{\rm yr}^{-1}$, and the population masses
in all cases nearly double.   

One of the important questions concerning the fate of the putatively
cooling gas in cooling flows is whether it is deposited in 
stars.  In Section 4 we found the cooling rate to low temperatures
to be $ 26 < \dot{M} < 42 ~M_{\odot}~{\rm yr}^{-1}$ within a radius of
137 arcsec.
After accounting for dust, and assuming that all of the blue light
in the Abell 2052 central galaxy is emerging from stars,
the star formation rate associated with constant star formation
over 1 Gyr is $0.6~M_{\odot}~{\rm yr}^{-1}$ within a radius of 3 arcsec.
We cannot directly compare the cooling in the X-ray gas with the star 
formation in the same $3\arcsec$ radius region because the AGN in the center 
of the cluster occupies the inner $2\farcs5$ region, and emits strongly in
the X-ray.  However, from the deprojected spectra described in \S4.2, we can 
measure the
cooling within the innermost annulus by modeling the emission with a MKCFLOW
component.  The innermost annulus has inner and outer radii of $2\farcs5$ and
$20\farcs7$ arcsec, respectively.  This is the closest to the center of the
cluster we can get with the deprojection while meeting our requirement that 
at least
$25\%$ of the emission is coming from the annulus of interest, and not from
projected emission, thus enabling us to constrain the physical parameters of
the gas.  For the cooling flow fit in this inner region, we measure
$\dot{M} = 12\pm1~ M_{\odot}~{\rm yr}^{-1}$.  This value would
almost certainly be lower at smaller radii and may be consistent with the
star formation rate if it could be measured from the 3 arcsec radius region.
For the aging,
instantaneous burst star formation scenarios, which consume their fuel in a 
very short period of time, and using the extinction corrected masses of 
$M_{\rm AP}\sim 1.8\times 10^8~ {\rm M}_\odot - 2.4\times 10^9~ {\rm M}_\odot$
derived above, the observed accretion population 
masses would correspond to a burst of star formation
following a buildup of material accreted from
the cooling flow  that occurred on timescales of
between 15 and 200 Myr for cooling at $12~{\rm M}_\odot ~{\rm yr}^{-1}$.
To within the uncertainty of the data, the color profile of the galaxy 
beyond the inner $3\arcsec$ is consistent with that of a typical cD galaxy
(McNamara \& O'Connell 1992).   Therefore, there is no evidence
for additional star formation at larger radii that would be occurring with
a Salpeter-like initial mass function.

Our analysis and conclusions
differ markedly from those of Martel et al.\ (2002) who argue
for a vastly smaller star formation rate. Therefore, a comment
on our respective analysis seems appropriate.  
The discrepancy between our respective star formation rates is due to
Martel's focus on the small blue filament, which is indeed
young and very low mass.  However, this filament emits
only a tiny fraction of the total excess blue light in
the inner 3 kpc of the galaxy, and is a negligible component
of the color excess seen in ground observations 
(McNamara \& O'Connell 1989, 1992; Crawford et al.\ 1999).
While the extended blue component seen from the ground is
also seen in the {\it Hubble} images, Martel et al.\ attribute this 
``optical bump''  to the so-called UV upturn seen in
normal elliptical galaxies.  Their interpretation seems
to us unlikely, as the central optical colors of the 
Abell 2052 galaxy are bluer than those of a normal giant elliptical galaxy.  
Our star formation rates are in much better agreement with
those found by Crawford et al.\ (1999) and McNamara \&
O'Connell (1989) when aperture and dust corrections
are taken into account.

In summary, using newer data, the substantially revised 
cooling and star formation rates
found here are now much closer to being in line.
While  this development bodes well for the hypothesis that 
star formation is being fueled by the cooling flow, 
cooling and star formation are occurring at substantially smaller
rates than had been reported in the past, suggesting that
energy is being fed back into the hot gas and preventing cooling
at the high rates reported from previous X-ray observatories. 

\section{Can Radio Source Heating Balance the Cooling Flow?}

As we have shown in Paper I and Soker, Blanton, \& Sarazin (2002), the
{\it Chandra} data are inconsistent with the bright X-ray shells being 
formed by a strong shock.  However, it is likely that a weak shock occurred
early in the life of the radio source.  Even in the absence of strong 
shock heating, we can test whether the energy output of the radio source is
sufficient to offset the energy put into the center of the cluster by the
cooling flow.  In other words, is the radio source capable of heating 
the amount of gas that we measure to be cooling into the cluster center from
the cooling flow?

The luminosity of isobaric cooling gas is given by
\begin{equation}
L_{\rm{cool}} = \frac{5}{2}\frac{kT}{\mu m_{p}} \dot{M} ~\rm{erg~s^{-1}}.
\end{equation}
Using a temperature of $kT = 3$ keV, and our upper limit on the 
mass-deposition rate of $42~M_{\odot}$ yr$^{-1}$, we find
$L_{\rm{cool}} \le 3.2 \times 10^{43}$ erg s$^{-1}$.

In Paper I, we calculated the energy output of the radio source by assuming
that the pressure within the radio bubbles was equal to that measured in
the shells of surrounding X-ray bright gas, or $P \approx 1.5 \times
10^{-10}$ dyn cm$^{-2}$.  Using the inner shell radii given in 
\S\ref{sec:shmass} as the bubble radii for the the northern and southern
bubbles, we
find that the total energy output of the radio source including the work
done on compressing the intracluster gas is 
$E_{\rm{radio}} \approx $ 5/2~PV $\approx 10^{59}$ ergs, where the factor
of 5/2 assumes that most of the energy within the bubbles is due to
nonrelativistic thermal plasma.  The factor is $\sim 2$ for any pressure
source.
As has been evidenced by the ``ghost cavities'' (radio-faint holes in the 
X-ray gas away from the cluster centers) in the Perseus (Fabian et al.\ 2000;
Churazov et al.\ 2000) 
and Abell~2597 clusters (McNamara et al.\ 2002), radio activity in cooling flow
clusters is episodic with a repetition rate of
$t_{\rm{rep}} \approx 10^8$ yr.  If the radio activity has a similar repetition
rate in Abell~2052, the average rate of energy output from the central radio 
source over the lifetime of the cluster is 
$E_{\rm{radio}}/t_{\rm{rep}} = 3.2 \times 10^{43}$ erg s$^{-1}$, 
which is approximately sufficient to offset the cooling rate, although all of
this energy will not go into heating the cooling gas.

\section{Conclusions}

We have presented a detailed analysis of the {\it Chandra} ACIS-S3 observation
of Abell 2052, including the large-scale cooling flow properties and
the central radio source / X-ray gas interaction region.  We found an average
cluster temperature of 2.9 keV, and an abundance of 0.6 times the
solar value.  The mass of gas
cooling to very low temperatures 
is  $26 < \dot{M} < 42 \, M_{\odot}$ yr$^{-1}$.
This value is 
approximately a factor of three lower than previous measurements from
{\it Einstein}, {\it ROSAT}, and {\it ASCA}.

We extracted spectra from seventeen circular annuli with average radii 
ranging from $12\arcsec$ to
$322\arcsec$.  The spectra were initially fitted with single-temperature
models.  We also performed a spectral deprojection to more accurately 
determine the values of temperature and abundance as a function of radius.
There is a sharp decline of the temperature with radius towards the center
of the cluster, with an average outer temperature of 3 keV dropping to
approximately 1 keV at the cluster center.  The abundance is approximately 
0.3 times
solar in the outer regions of the cluster, rises to the solar 
value between 30 and 40 arcsec from the cluster center (exterior to the 
bright, X-ray shells), and drops again to 0.4 times solar at the center.
One explanation for this abundance profile 
is that the ICM is inhomogeneous on small scales ($\la 1$ kpc),
containing mostly very metal poor gas along with clumps of metal rich
gas
(Fabian et al.\ 2001; Morris \& Fabian 2001).
Another possibility is that resonance scattering redistributes the emission
from strong emission lines.
However, there are problems with both scenarios (B\"ohringer et al.\ 2001,
Sakelliou et al.\ 2002).

Emission from the central source, 3C 317, was well described as a power-law
with a photon index of $\Gamma = 2.0$, and Galactic absorption.  There is
no evidence for excess absorption,
with the upper limit being $4.4 \times 10^{20}$ cm$^{-2}$.
The AGNs
luminosity in the 0.7 -- 7.0 keV band is $L_X = 7.9 \times 10^{41}$ erg 
s$^{-1}$.

Gas and total mass profiles were determined.  The total mass within 
$261\arcsec$ (248 kpc) is $4.0 \times 10^{13}~M_{\odot}$.  The gas mass
fraction increases from approximately $5\%$ in the inner regions of the
cluster to approximately $10\%$ at the outermost region sampled.

Results on the radio source / X-ray gas interaction region at the
center of Abell 2052 were presented (see Paper I for more discussion of
this region). 
Temperature and abundance maps show that the brightest parts of the X-ray
shells have the lowest temperatures and abundances (in projection) in the
cluster.  Again, this is consistent with what would be expected in a 
chemically inhomogeneous
ICM (Fabian et al.\ 2001; Morris \& Fabian 2001), where metal-rich gas
cools quickly below the X-ray regime by line emission, leaving behind cool,
metal-poor gas.  This would explain the lack of line emission, 
that would be expected if assuming a homogeneous ICM, measured
from low-temperature gas (Peterson et al.\ 2001) 

The total mass for the two bright shells was updated to a value of
$5 \times 10^{10} M_{\odot}$.  This mass is consistent with the gas
in the shells being compressed out of the X-ray holes by the radio source,
and with the holes being devoid of X-ray emitting gas.  

The pressure derived
from the X-ray observations in the bright shells is about an order of 
magnitude higher than that derived using radio observations and assuming
equipartition (Zhao et al.\ 1993).
We explored the possibility that the extra pressure needed to support the
shells comes from very hot, diffuse, thermal gas filling the X-ray holes.
Using radio observations including
the Faraday rotation measure, and the lack of Faraday depolarization (Ge \& 
Owen 1994) seen in the region of the X-ray holes, we derived a lower limit
on the temperature of a hot, diffuse component of $kT \ga 20$ keV
if this gas provides the missing pressure support for the radio
bubbles.
In addition, we calculated the magnetic field in the shells based on the
Faraday rotation measure, and found $B \approx 11~\mu$G.
The magnetic pressure was found to be 30 times lower than the gas pressure.

Based on recent $HST$ results (Martel et al.\ 2002) and a reanalysis 
of U-band imaging and spectroscopy (McNamara \& O'Connell 1989, 1992),
we determined the star formation rate at the center of Abell 2052 and
found a 
value of $0.6~M_{\odot}$ yr$^{-1}$ for the inner 3\arcsec.
From the {\it Chandra} data, within a radius of approximately 20\arcsec\
of the cluster center (and excluding the emission from the AGN), we measured a
mass-deposition rate of $12\pm1~M_{\odot}$ yr$^{-1}$.  We cannot
measure the X-ray cooling rate within the inner 3\arcsec\ to compare 
directly with the star formation rate because non-thermal X-ray emission 
from the AGN  is the dominant source of photons in this region.  However, it
is almost certain that the mass-deposition rate within 3\arcsec\ is less than
$12~M_{\odot}$ yr$^{-1}$ and may very well be consistent with the star
formation rate within that region.

Finally, although evidence for strong shocks or heating from the radio
source is not seen, we find that the 
total combined energy output
of multiple outbursts of the radio source, occurring episodically with a
repetition period of $t_{\rm{rep}} \sim 10^8$ yr over the lifetime of
the cluster, can offset the cooling from the cooling flow if this energy
can be used to effectively heat the cooling gas.

\acknowledgements
We thank Noam Soker for useful discussions.
Support for this work was provided by the National Aeronautics and Space
Administration through {\it Chandra} Award Numbers
GO0-1158X
and
GO1-2133X,
issued by the {\it Chandra} X-ray Center,
which is operated by the Smithsonian Astrophysical Observatory for and on
behalf of NASA under contract NAS8-39073.
Support for E. L. B. was provided by NASA through the {\it Chandra}
Fellowship
Program, grant award number PF1-20017, under NASA contract number
NAS8-39073.
B. R. M. acknowledges generous support from 
NASA Long Term Space Astrophysics grant NAG5-11025, Chandra General
Observer Program award AR2-3007X, and grant 47735-001-023N from
the Department of Energy through the Los Alamos National Laboratory.

\begin{deluxetable}{ccccccc}
\tabletypesize\footnotesize
\tablewidth{0pt}
\tablecaption{Fits to the Total Spectrum \label{tab:tot}}
\tablehead{
\colhead{Model} & \colhead{$N_H$} & \colhead{$kT_{\rm low}$} & 
\colhead{$kT_{\rm high}$} & \colhead{Abund} & \colhead{$\dot{M}$} & 
\colhead{$\chi^{2}$/d.o.f.} \\
\colhead{} & \colhead{($10^{20}$ cm$^{-2}$)} & \colhead{(keV)} 
& \colhead{(keV)} & \colhead{(solar)} & \colhead{($M_{\odot}$/yr)} 
& \colhead{}}  
\startdata
1-MEKAL & (2.85) & \nodata & 2.72$^{+0.035}_{-0.025}$ & 
0.53$^{+0.024}_{-0.028}$ & \nodata & 676.3/475$ = $1.42 \\
1-MEKAL-abs & 0$^{+0.24}_{-0}$ & \nodata & 2.86$^{+0.029}_{-0.025}$ & 
0.60$^{+0.022}_{-0.038}$ & \nodata & 599.9/474$ = $1.27 \\
2-MEKAL & (2.85) & 0.77$^{+0.065}_{-0.053}$ & 2.87$^{+0.038}_{-0.022}$ & 
0.64$^{+0.042}_{-0.032}$ & \nodata & 560.9/473$ = $1.19 \\
2-MEKAL-abs & 2.34$^{+0.97}_{-0.40}$ & 0.81$^{+0.068}_{-0.048}$ & 
2.90$^{+0.040}_{-0.028}$ & 
0.66$^{+0.025}_{-0.043}$ & \nodata & 560.8/472$ = $1.19 \\
MKCFLOW+MEKAL & (2.85) & 0.43$^{+0.12}_{-0.43}$ & 2.95$^{+0.069}_{-0.021}$ & 
0.64$^{+0.027}_{-0.047}$ & 32$^{+4}_{-4}$ &  554.6/473$ = $1.17 \\
MKCFLOW+MEKAL & (2.85) & (0.001) & 2.94$^{+0.060}_{-0.026}$ & 
0.64$^{+0.033}_{-0.041}$ & 28$^{+4}_{-2}$ &  554.9/474$ = $1.17 \\
MKCFLOW+MEKAL-abs1 & 3.81$^{+1.06}_{-0.97}$ & 0.01$^{+0.48}_{-0.009}$ & 
2.95$^{+0.069}_{-0.026}$ & 
0.64$^{+0.025}_{-0.043}$ & 37$^{+5}_{-7}$ & 553.8/472$ = $1.17 \\
MKCFLOW+MEKAL-abs2 & 3.83$^{+1.04}_{-0.98}$ & (0.001) & 
2.95$^{+0.069}_{-0.025}$ & 
0.64$^{+0.025}_{-0.043}$ & 37$^{+5}_{-7}$ & 553.8/473$ = $1.17 \\
\enddata
\end{deluxetable}

\begin{deluxetable}{crccccccc}
\tabletypesize\footnotesize
\tablewidth{0pt}
\tablecaption{Temperature and Abundance Profiles \label{tab:kT}}
\tablehead{
\colhead{} & \colhead{} & \multicolumn{3}{c}{Projected} & &
\multicolumn{3}{c}{Deprojected} \\
\cline{3-5} \cline{7-9}
\colhead{} & \colhead{$r$} & \colhead{$kT$} & 
\colhead{Abund} & 
\colhead{} & \colhead{} &
\colhead{$kT$} & 
\colhead{Abund} & 
\colhead{} \\
\colhead{Annulus} & \colhead{(\arcsec)} & \colhead{(keV)} &
\colhead{(solar)} & \colhead{$\chi^{2}$/d.o.f.} & \colhead{} & \colhead{(keV)} 
& \colhead{(solar)} & \colhead{$\chi^{2}$/d.o.f.}  }  
\startdata
 1 & $2.5-20.7$ & 1.62$^{+0.031}_{-0.033}$ & 0.39$^{+0.039}_{-0.036}$ & 390.4/157$ = $2.49 && 1.10$^{+0.028}_{-0.029}$ & 0.39$^{+0.13}_{-0.10}$ & 245.5/157$ = $1.56\\
 2 & $20.7-24.6$ & 1.78$^{+0.041}_{-0.044}$ & 0.46$^{+0.074}_{-0.067}$ & 206.9/115$ = $1.80 && 1.31$^{+0.080}_{-0.12}$ & 0.32$^{+0.18}_{-0.13}$ & 181.9/115$ = $1.58\\
 3 & $24.6-30.5$ & 2.25$^{+0.058}_{-0.063}$ & 0.69$^{+0.098}_{-0.088}$ & 155.7/139$ = $1.12 && 1.71$^{+0.10}_{-0.12}$ & 0.52$^{+0.21}_{-0.16}$ & 147.6/139$ = $1.06\\
 4 & $30.5-37.4$ & 2.73$^{+0.090}_{-0.094}$ & 0.83$^{+0.12}_{-0.11}$ & 176.8/154$ = $1.15 && 2.32$^{+0.22}_{-0.14}$ & 1.15$^{+0.39}_{-0.34}$ & 178.6/154$ = $1.16\\
 5 & $37.4-42.3$ & 3.01$^{+0.17}_{-0.14}$ & 0.71$^{+0.13}_{-0.12}$ & 143.4/138$ = $1.04 && 3.51$^{+0.49}_{-0.44}$ & 1.03$^{+0.57}_{-0.42}$ & 144.2/138$ = $1.04\\
 6 & $42.3-48.2$ & 2.75$^{+0.11}_{-0.12}$ & 0.57$^{+0.11}_{-0.10}$ & 146.4/137$ = $1.07 && 2.45$^{+0.38}_{-0.21}$ & 0.54$^{+0.42}_{-0.21}$ & 149.0/137$ = $1.09\\
 7 & $48.2-54.1$ & 2.88$^{+0.15}_{-0.13}$ & 0.60$^{+0.14}_{-0.12}$ & 164.5/136$ = $1.21 && 2.31$^{+0.47}_{-0.35}$ & 0.43$^{+0.46}_{-0.24}$ & 165.0/136$ = $1.21\\
 8 & $54.1-64.0$ & 3.14$^{+0.14}_{-0.15}$ & 0.64$^{+0.11}_{-0.10}$ & 170.8/170$ = $1.00 && 3.15$^{+0.37}_{-0.35}$ & 1.16$^{+0.54}_{-0.39}$ & 170.6/170$ = $1.00\\
 9 & $64.0-76.3$ & 3.14$^{+0.14}_{-0.15}$ & 0.45$^{+0.085}_{-0.077}$ & 176.1/189$ = $0.93 && 3.20$^{+0.55}_{-0.39}$ & 0.58$^{+0.52}_{-0.23}$ & 177.1/189$ = $0.94\\
10 & $76.3-91.0$ & 3.14$^{+0.13}_{-0.13}$ & 0.38$^{+0.072}_{-0.067}$ & 214.7/206$ = $1.04 && 2.74$^{+0.37}_{-0.32}$ & 0.30$^{+0.20}_{-0.15}$ & 215.1/206$ = $1.04\\
11 & $91.0-108.2$ & 3.29$^{+0.14}_{-0.14}$ & 0.40$^{+0.080}_{-0.073}$ & 242.4/221$ = $1.10 && 4.23$^{+0.53}_{-0.54}$ & 0.74$^{+0.40}_{-0.32}$ & 240.9/221$ = $1.09\\
12 & $108.2-127.9$ & 2.91$^{+0.15}_{-0.11}$ & 0.32$^{+0.069}_{-0.060}$ & 232.9/231$ = $1.01 && 2.58$^{+0.30}_{-0.30}$ & 0.26$^{+0.17}_{-0.12}$ & 231.4/231$ = $1.00\\
13 & $127.9-152.5$ & 3.13$^{+0.16}_{-0.16}$ & 0.35$^{+0.080}_{-0.072}$ & 215.5/215$ = $1.00 && 3.04$^{+0.43}_{-0.36}$ & 0.41$^{+0.25}_{-0.19}$ & 215.4/215$ = $1.00\\
14 & $152.5-187.0$ & 3.21$^{+0.15}_{-0.15}$ & 0.31$^{+0.070}_{-0.065}$ & 256.9/259$ = $0.99 && 3.23$^{+0.36}_{-0.35}$ & 0.31$^{+0.18}_{-0.15}$ & 257.0/259$ = $0.99\\
15 & $187.0-236.2$ & 3.19$^{+0.15}_{-0.15}$ & 0.33$^{+0.069}_{-0.063}$ & 334.3/320$ = $1.04 && 3.52$^{+0.42}_{-0.36}$ & 0.25$^{+0.18}_{-0.15}$ & 333.9/320$ = $1.04\\
16 & $236.2-285.4$ & 2.89$^{+0.21}_{-0.18}$ & 0.34$^{+0.11}_{-0.10}$ & 231.7/221$ = $1.05 && 3.39$^{+0.58}_{-0.55}$ & 0.71$^{+0.29}_{-0.34}$ & 230.8/221$ = $1.04\\
17 & $285.4-359.2$ & 2.65$^{+0.22}_{-0.22}$ & 0.19$^{+0.11}_{-0.089}$ & 261.2/230$ = $1.14 && 2.65$^{+0.22}_{-0.22}$ & 0.19$^{+0.11}_{-0.089}$ & 261.2/230$ = $1.14\\
\enddata

\end{deluxetable}

\begin{deluxetable}{crcc}
\tabletypesize\footnotesize
\tablewidth{0pt}
\tablecaption{Absorption Profile \label{tab:nh}}
\tablehead{
\colhead{} & \colhead{$r$} & \colhead{$N_{H}$(1T)} &
\colhead{} \\
\colhead{Annulus} & \colhead{(\arcsec)} &  \colhead{($10^{20}$ cm$^{-2}$)} 
& \colhead{$\chi^{2}$/d.o.f.} }
\startdata
1 &  $2.5-20.7$    & $0.00^{+0.28}_{-0.00}$ & 355.9/233$ = $2.28 \\
2 &  $20.7-24.6$   & $0.00^{+0.36}_{-0.00}$ & 187.0/114$ = $1.64  \\
3 &  $24.6-30.5$   & $0.00^{+1.10}_{-0.00}$ & 147.7/138$ = $1.07  \\
4 &  $30.5-37.4$   & $1.95^{+1.90}_{-1.86}$ & 176.2/153$ = $1.15  \\
5 &  $37.4-42.3$   & $1.45^{+2.12}_{-1.45}$ & 142.2/137$ = $1.04  \\
6 &  $42.3-48.2$   & $3.10^{+2.37}_{-2.34}$ & 146.3/136$ = $1.08  \\
7 &  $48.2-54.1$   & $0.14^{+2.20}_{-0.14}$ & 160.4/135$ = $1.19  \\
8 &  $54.1-64.0$   & $0.87^{+1.81}_{-0.87}$ & 167.6/169$ = $0.99  \\
9 &  $64.0-76.3$   & $1.97^{+1.77}_{-1.75}$ & 175.4/188$ = $0.93  \\
10 & $76.3-91.0$   & $1.14^{+1.63}_{-1.14}$ & 211.7/205$ = $1.03  \\
11 & $91.0-108.2$  & $0.00^{+1.00}_{-0.00}$ & 229.7/220$ = $1.04  \\
12 & $108.2-127.9$ & $0.00^{+0.89}_{-0.00}$ & 218.4/231$ = $0.95  \\
13 & $127.9-152.5$ & $0.35^{+1.83}_{-0.35}$ & 210.5/214$ = $0.98  \\
14 & $152.5-187.0$ & $0.80^{+1.68}_{-0.80}$ & 252.8/258$ = $0.98  \\
15 & $187.0-236.2$ & $0.00^{+0.71}_{-0.00}$ & 317.1/319$ = $0.99  \\
16 & $236.2-285.4$ & $0.00^{+1.07}_{-0.00}$ & 222.3/220$ = $1.01  \\
17 & $285.4-359.2$ & $0.00^{+1.52}_{-0.00}$ & 255.0/229$ = $1.11  \\
\enddata

\end{deluxetable}

\end{document}